\documentclass[prd,showpacs,preprint,tightenlines,floatfix,preprintnumbers,nofootinbib,eqsecnum,draft]{revtex4}
 \usepackage{color}

 \usepackage[dvips,final]{graphicx}
  \usepackage{amssymb,amsmath,bm,pifont}

\newcommand{\gev}[1]{\relax\ifmmode{\text{GeV}^{#1}}              
                     \else{GeV$^{#1}${ }}\fi}                     
\newcommand{\Ds}{\displaystyle}
\newcommand{\va}[1]{\langle{#1}\rangle}
\def\muF{\relax\ifmmode\mu_\text{F}^2\else{$\mu_\text{F}^2${ }}\fi}
\def\muR{\relax\ifmmode\mu_\text{R}^2\else{$\mu_\text{R}^2${ }}\fi}
\def\muO{\relax\ifmmode{\mu_{0}^{2}}\else{$\mu_{0}^{2}${ }}\fi}

\begin{document}
\thispagestyle{empty}
\date{\today}
\preprint{\hbox{RUB-TPII-03/09}}
\title{QCD sum rules with nonlocal condensates and the spacelike
       pion form factor \\ }
\author{A.~P.~Bakulev}
 \email{bakulev@theor.jinr.ru}
  \affiliation{Bogoliubov Laboratory of Theoretical Physics,
               JINR, 141980 Dubna, Moscow region, Russia\\ }

\author{A.~V.~Pimikov}
 \email{pimikov@theor.jinr.ru}
  \affiliation{Bogoliubov Laboratory of Theoretical Physics,
               JINR, 141980 Dubna, Moscow region, Russia\\ }

\author{N.~G.~Stefanis}
 \email{stefanis@tp2.ruhr-uni-bochum.de}
  \affiliation{Institut f\"{u}r Theoretische Physik II,
               Ruhr-Universit\"{a}t Bochum, D-44780 Bochum, Germany\\}

\vspace {10mm}
\begin{abstract}
We present a detailed investigation of the spacelike pion's
electromagnetic form factor using three-point QCD sum rules that
exclusively involve nonlocal condensates.
Our main methodological tools are a spectral density which includes
$O(\alpha_s)$ corrections, a suitably improved
Gaussian ansatz to model the distribution of the average momentum
of quarks in the QCD vacuum, and a perturbative scheme that avoids
Landau singularities.
Using this framework, we obtain predictions for the pion form factor
together with error estimates originating from intrinsic theoretical
uncertainties owing to the perturbative expansion and the
nonperturbative method applied.
We also discuss our results in comparison with other calculations,
in particular, with those following from  the AdS/QCD correspondence.
We find good agreement of our predictions with measurements in the
range of momenta covered by the existing experimental data between
$1-10$~GeV$^2$.

\end{abstract}
\maketitle

\section{Introduction}
\label{sec:intro}

The description of hadron form factors within QCD represents a major
challenge because in such exclusive processes intact hadrons appear
in the initial and final states, bearing the quark-gluon
binding effects controlled by nonperturbative dynamics.
Essential to implementing perturbation theory is the factorization
assumption of a short-distance part from a large-distance remainder,
the latter being mainly nonperturbative.
It was proven long ago \cite{ER80,ER80tmf,LB79,LB80} that the pion's
electromagnetic form factor can be factorized at large $Q^2$ in the
form of a convolution of a hard-scattering amplitude $T_\text{H}$ and
two distribution amplitudes (DA) $\varphi_\pi$ \cite{Rad77} to describe
the initial and final pion states.

Following this type of approach, it has been shown by many authors
(see, e.g., \cite{Rad84,IL84,JK93,SSK00,BPSS04} and references cited
therein) that the factorizable (hard) part of the electromagnetic pion
form factor is too small to afford for a good agreement with the
available experimental data \cite{FFPI78,JLab08II}.
Admittedly, the quality of the existing data at higher $Q^2$ values is
too poor to draw any definitive conclusions, though---at least the
tendency---is clear: if at al, the perturbative regime seems to
be far outside the currently probed momentum scales.
For that reason, it seems reasonable to pursue alternative methods to
compute the pion's electromagnetic form factor which do not rely on
perturbation theory.

Our present work is in the context of a three-point AAV
(A for axial, V for vector current) correlator, having recourse to
the operator product expansion (OPE) and a dispersion relation.
Our techniques are closely related to the QCD sum-rule approach of
\cite{BR91,Rad95}, but with the important difference that we use
exclusively nonlocal condensates.
In particular, a quark-gluon-antiquark nonlocal condensate is employed
in which all three inter-parton separations are nonlocal.

Moreover, we use a spectral density which includes terms of
$O(\alpha_s)$.
The influence of this next-to-leading-order (NLO) contribution to
the pion form factor turns out to be quite important, reaching the
level of $20\%$.
To describe the momentum distribution of vacuum quarks, we go beyond
the minimal Gaussian model, used before \cite{MR92,BM98,BMS01},
and consider also a model, obtained more recently by two of us
\cite{BP06}, that has the following methodological advantages:
(a) it satisfies the QCD equations of motion and (b) it minimizes
the non-transversality of the VV correlator.

The principal results of this paper are the following:\\
(i)   We give a general formalism for the calculation of the pion's
      electromagnetic form factor that contains several methodological
      improvements relative to all previous approaches employing QCD
      sum rules with vacuum (local and nonlocal) condensates.\\
(ii)  We suggest a way how to determine the threshold in the
      local-duality approach for intermediate values of $Q^2$
      where measurements have already been carried out or are
      planned.\\
(iii) We present predictions for the pion form factor, including also
      inherent theoretical uncertainties, that are in good agreement
      with new lattice results and real experimental data from the
      Cornell and JLab Collaborations.

The plan of this paper is as follows.
In the next section we recall the standard QCD description of the
pion's electromagnetic form factor according to the factorization
theorem.
A brief report on the current status of the obtained predictions
is given, avoiding technicalities.
In Sec.\ \ref{sec:vac}, we discuss the method of QCD sum
rules with local and nonlocal condensates in the calculation of
the pion form factor.
The analysis of the present work is detailed in Sec.\
\ref{sec:pion-FF-analysis}.
We present our computations and the obtained results and compare
them with other theoretical predictions, recent lattice simulations,
and experimental data.
This section contains also estimates of the inherent theoretical
uncertainties of the applied method.
In Sec.\ \ref{sec:Conclusions} we state our conclusions and summarize
our main results.

\section{Convolution scheme for the pion form factor in QCD}
\label{sec:conv}
Applying the factorization theorem \cite{ER80tmf,LB80}, the pion's
electromagnetic form factor can be written in the form (for reviews
see, for instance, \cite{CZ84,BL89,Ste99})
\begin{eqnarray}
    F_{\pi}(Q^2)
 &=& \int\limits_{0}^{1}\!\!\!\int\limits_{0}^{1}\!
      \varphi_{\pi}(x,\mu^2)\,
       T_\text{H}(x,y;Q^2,\mu^2)\,
        \varphi_{\pi}(y,\mu^2)\,
         dx\,dy\,.
 \label{eq:F-fact}
\end{eqnarray}
The hard amplitude is the sum of all Feynman diagrams in which the
struck quark is connected to the spectator via highly off-shell gluon
propagators, meaning that the transverse interquark distance is
rather small, i.e., of the order of the inverse large momentum
transfer $Q$, and that both partons share comparable fractions of
longitudinal momentum $x_i=p_i^+/P^+$, where the parton four-momentum
in light-cone coordinates is
$p_i=(p_i^+, p_i^-,\mathbf{k}_{\perp i})$ and
$P$ is the initial four-momentum of the pion.
Momentum conservation under the assumption of exact $u-d$ symmetry
implies that
$\sum_{i=1}^{2}x_i=1$ and $\sum_{i=1}^{2}\mathbf{k}_{\perp i}=0$.
In our case, $x_1=x$ and $x_2=1-x\equiv \bar x$.

The unknown binding effects of the pion state have been absorbed into
the pion DA which at the leading-twist level two is represented by the
valence-state wave function on the light cone averaged over transverse
momenta up to the factorization scale $\mu$ of the process.
It is defined by the following universal operator matrix element
(see, e.g., \cite{CZ84} for a review)
\begin{eqnarray}
  \va{0\mid\bar{d}(z)\gamma^{\mu}\gamma_5\,
 {\mathcal C}(z,0) u(0)\mid\pi(P)}
  \Big|_{z^2=0}
 &=& i f_{\pi} P^{\mu}
       \int^1_0 dx e^{ix(zP)}\
      \varphi_{\pi}\left(x,\muO\right)
\label{eq:pi-DA-fpi}
\end{eqnarray}
with the normalization
\begin{equation}
 \int_0^1 \varphi_{\pi}(x,\muO)\, dx
  = 1 \ ,
\label{eq:norm}
\end{equation}
where $f_{\pi} = 130.7 \pm 0.4$~MeV \cite{PDG08} is the pion decay
constant defined by
\begin{equation}
\langle 0|\bar{d}(0)\gamma_{\mu} \gamma_{5} u(0)|\pi^{+}(P) \rangle
 =
iP_{\mu} f_{\pi} \ .
\label{eq:fpi}
\end{equation}
Here
\begin{equation}
  {\mathcal C}(z,0)
  = {\mathcal P}
  \exp\!\left[-ig_s\!\!\int_0^z t^{a} A_\mu^{a}(y)dy^\mu\right]
\label{eq:pexponent}
\end{equation}
is the Fock--Schwinger phase factor
(termed in \cite{Ste84} the color ``connector''), path-ordered along
the straight line joining the points $0$ and $z$, to preserve gauge
invariance.
Note that the scale $\muO$, called the normalization scale of the
pion DA, is related to the ultraviolet (UV) regularization of the
quark-field operators on the light cone whose product becomes singular
for $z^2=0$.
The derivation of hadron distribution amplitudes at finite momentum
transfer is outside perturbative QCD; it requires nonperturbative
approaches, like QCD sum rules with vacuum condensates \cite{CZ84},
or lattice calculations.
The latter method has recently reached a high level of precision
in calculating the first and second moment of the pion DA
\cite{Lat05,Lat06,Lat07}, though the calculation of the fourth moment
is still pending, while recently a prediction for its value was
worked out \cite{Ste08} by combining the CLEO data \cite{CLEO98} and
the lattice calculation for the second moment.
Here, an alternative method proposed by Braun and M\"uller
\cite{BraM07} may be proven useful.
On the other hand, QCD sum rules incorporate information about
the non-trivial structure of the QCD vacuum and, therefore, constitute
a useful analytic tool for determining hadron distribution amplitudes
using local constraints on their moments (inverse-scattering method).
Chernyak and Zhitnitsky \cite{CZ84} extracted the lowest few moments
$
 \langle \xi ^N\rangle
\equiv
 \int_{0}^{1} \varphi_{\pi}(x)(2x-1)^N\ dx
$
of $\varphi_\pi$ using correlators of two axial currents with local
operators containing $N$ covariant derivatives \cite{Rad77,ER80}.
However, it was pointed out later \cite{MR86,MR89,MR92} that QCD sum
rules with local condensates strongly emphasize the endpoint region,
because the condensate terms are strongly peaked just at these
endpoints $x=0,1$, where one of the quarks has a vanishing virtuality.
Therefore, the authors of Refs.\ \cite{MR86,MR89,MR92} suggested
to introduce nonlocal condensates that can account for the
possibility that vacuum quarks may have a finite virtuality
(a nonzero average transverse momentum).
As a result, Feynman-type configurations, in which one of the quark
carries almost the entire available momentum while the spectators are
``wee'' (with almost vanishing virtualities), are separated out from
the pion DA and treated separately in a non-factorizing (soft)
contribution to the form factor (for a discussion of the Feynman
mechanism in this context, see \cite{SSK99}).

\begin{figure}[t]
 \centerline{\includegraphics[width=0.5\textwidth]{
  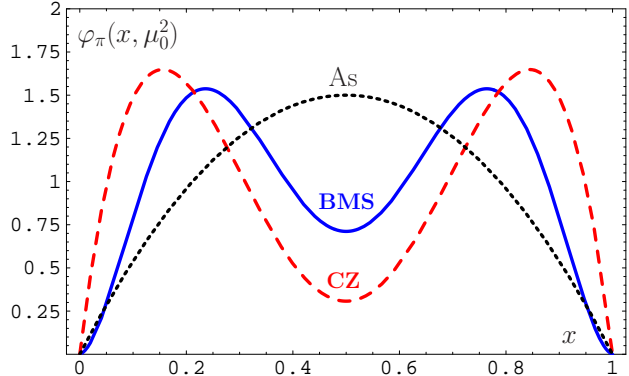}}
  \caption{\footnotesize Comparison of selected pion DAs labeled by
   obvious acronyms:
   $\varphi_\text{as}$ (dotted line),
   $\varphi_\text{CZ}$ (dashed line) \protect\cite{CZ84}, and
   $\varphi_\text{BMS}$ (solid line) \protect\cite{BMS01}, defined by
   the coefficients $a_2$ and $a_4$ in Table~\ref{tab:DAs}.
   All DAs are normalized at the same scale $\muO\approx1$ GeV$^2$.
   \label{fig:pion_das}}
\end{figure}

More recently, a QCD sum-rule analysis was carried out \cite{BMS01}
which employs nonlocal condensates parameterized for simplicity in
terms of a single mass-scale parameter
$\lambda_{q}^{2}=\langle k_{\perp}^2\rangle$
with values in the range \cite{BI82lam,OPiv88,DDM99,BM02}
\begin{eqnarray}
  \lambda_q^2
=
  0.35-0.55~\text{GeV}^{2}\, .
\label{eq:lambda.q.SR}
\end{eqnarray}
\begin{table}[b]\vspace*{-3mm}
\begin{ruledtabular}
\begin{tabular}{c|cccl}
 ~~DA~~&~$a_2(1~\text{GeV}^2)$~
                  &~$a_4(1~\text{GeV}^2)$~
                              &~$I_{-1}^{\pi}(1~\text{GeV}^2)$~
                                           &~Shape~~
\\ \hline
~~Asy~~&~~$0.00$~~&~~$~0.00$~~&~~ $3.00$ ~~&~convex~~
\\
~~CZ~~ &~~$0.56$~~&~~$~0.00$~~&~~ $4.68$ ~~&~double-humped, end-point enhanced~~
\\
~~BMS~~&~~$0.20$~~&~~$-0.14$~~&~~ $3.18$ ~~&~double-humped, end-point suppressed~~
\end{tabular}
\end{ruledtabular}\vspace*{-3mm}
\caption{Main characteristics of three different DAs, abbreviated by
acronyms:
CZ stands for Chernyak--Zhitnitsky \cite{CZ84} and BMS for
Bakulev--Mikhailov--Stefanis \cite{BMS01}.
Here $a_2$ and $a_4$ are the second and fourth Gegenbauer coefficients,
respectively, whereas $I_{-1}^{\pi}$ denotes the inverse moment of
$\varphi_\pi$.
 \label{tab:DAs}
 }
\end{table}
This way, the first ten moments of $\varphi_\pi$ at the initial scale
$\mu_{0}^{2}=1.35$~GeV$^2$
were determined with the following values (theoretical errors in
parentheses) \cite{BMS01}
\begin{eqnarray}
  \langle \xi^2\rangle= 0.265(20), ~~~
  \langle \xi^4\rangle= 0.115(12), ~~~
  \langle \xi^N\rangle\approx 0 , (N=6,8,10) \ .
\label{eq:moments-BMS}
\end{eqnarray}
Recasting the pion DA in terms of the Gegenbauer polynomials
$C_{n}^{3/2}(2x-1)$, which are the one-loop eigenfunctions of the ERBL
kernel \cite{ER80tmf,LB80}, one finds
\begin{eqnarray}
 \varphi_{\pi}(x,\mu^2)
  = \varphi^\text{as}(x)
     \left[ 1
          + \sum\limits_{n\geq1}a_{2n}(\mu^2)\, C_{2n}^{3/2}(2x-1)
     \right],~~~
 I^\pi_{-1}(\mu^2)
  = 3\left[1 + \sum\limits_{n\geq1}a_{2n}(\mu^2)
       \right],~~~
 \label{eq:Gegen}
\end{eqnarray}
where the asymptotic pion DA has the form
\begin{eqnarray}
 \varphi^\text{as}(x)
  = 6\,x\,(1-x)\,.
\end{eqnarray}
The values of the Gegenbauer coefficients, encoding the
nonperturbative dynamics, of some characteristic pion DAs are given
in Table \ref{tab:DAs}, while the corresponding profiles are displayed
in Fig.\ \ref{fig:pion_das}.

It is worth noting that the rate of convergence
$\varphi_\pi(x,\mu^2)\to\varphi^\text{as}(x)$,
or $a_{2n}(Q^2)\to0$,
at $Q^2\to\infty$
is not fast and is determined in the one-loop approximation
by the logarithmic law
\begin{eqnarray}
 \label{eq:1loop.Evo}
  a_{2n}(\mu^2)
   &=&
   a_{2n}(\mu_0^2)
    \left[\frac{\alpha_{s}(\mu^2)}{\alpha_{s}(\mu_0^2)}
    \right]^{\gamma_{2n}^{(0)}}\,.
\end{eqnarray}
Here
$\gamma_{2}^{(0)}=0.62$, $\gamma_{4}^{(0)}=0.90$,
whereas all other anomalous dimensions
$\gamma_{2n}^{(0)}\geq 1$ for $n\geq3$.
Numerical estimates show that if one has at some typical initial scale
$\mu^2\thickapprox1$~GeV$^2$ a coefficient
$a_{2}(1~\text{GeV}^2)=0.25$,
as indicated by the CLEO data \cite{CLEO98} and recent lattice
simulations \cite{BMS05lat,BMPS07}, its value would become $3$ times
smaller only at the tremendous scale of $\mu^2\sim 75400$~GeV$^2$,
which is certainly outside the reach of any experimental measurement.
The situation at intermediate momentum transfers
$20~\text{GeV}^2 \geq Q^2\geq 4$~GeV$^2$
is more delicate.
In this region, factorization partially fails owing to a quite sizable
contribution of the soft part which is non-factorizable.
Hence, one has to calculate this part of the pion form factor
using either phenomenological models~\cite{IL84,JK86,IL89,JK93},
or by employing
some nonperturbative concepts, like the method of QCD sum rules
(SR)~\cite{NR84,IS83,BR91}.
Still another option is to apply the local quark-hadron
duality approach~\cite{NR82,Rad95,BRS00}.
On the other hand, at the one-loop level and at asymptotically large
$Q^2$, the pion form factor turns out to be \cite{CZS77,FJ79}
\begin{eqnarray}
 \label{eq:FF.pi.Fact}
  F^{\text{pert}}_{\pi}(Q^2)
   = \frac{8\pi \alpha_s(Q^2) f_{\pi}^2}{9\,Q^2}
      \left|I^\pi_{-1}(Q^2)\right|^2
  ~~~\text{with}~~~
  I^\pi_{-1}(Q^2)
   = \int\limits_0^1\frac{\varphi_{\pi}(x,Q^2)}{x}\,dx \,.
\end{eqnarray}
The precise value of $Q^2$ at which this perturbative expression
should start to prevail cannot be predicted (determined) accurately.
The estimates for the cross-over momentum scale range from
$100$~GeV$^2$ \cite{IL84,JK93,BLM07} down to values around
$20$~GeV$^2$ \cite{SSK99,SSK00,BPSS04}.
But even this latter relatively small momentum is hopelessly
far away from the capabilities of any operating or planned accelerator
facility.

\section{QCD vacuum and the pion form factor}
 \label{sec:vac}
In view of these facts, one may look for alternative methods to
calculate the pion form factor in the experimentally accessible
region of momentum transfer.
Even after the commissioned 12~GeV upgrade for
Continuous Electron Beam Accelerator Facility (CEBAF), the expected
high-precision experimental data will not be sufficient to enter the
perturbative regime.
Hence, instead of predicating to the elusive hope for still higher
energies in the future, we discuss an alternative approach, based
on three-point QCD SR to the pion form factor~\cite{IS83,NR84}.
One of the advantages of this technique is that the shape of the
pion DA is irrelevant, reducing this way the theoretical uncertainty.

\begin{figure}[h]
 \centerline{\includegraphics[width=1\textwidth]{
  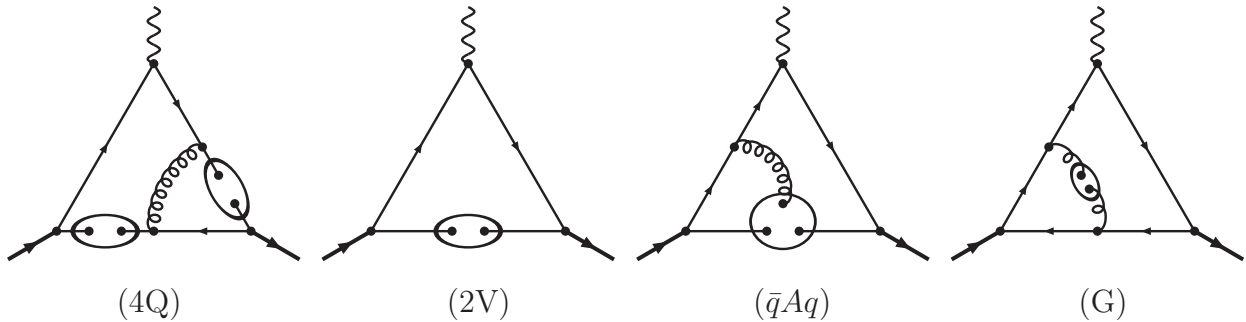}}
  \caption{\label{fig:JJJ.NLO}\footnotesize
   Nonperturbative contributions~\cite{NR82,IS82,BR91}
   to the QCD SR, (cf.\ Eq.\ (\ref{eq:ffQCDSR})).
   Here we show only typical for each subclass diagrams,
   whereas the complete set includes also mirror-conjugated diagrams
   (for the 4Q and $\bar{q}Aq$ subclasses) and
   diagrams with permutations of the gluon-lines insertions
   (subclass G).}
\end{figure}

The construction of the Borel SR for the pion form factor from the
three-point AAV correlator
\begin{equation}
  \int\!\!\!\!\int\!\!d^4x\,d^4y\,e^{i(qx-p_2y)}
  \langle 0|T\!\left[J^{+}_{5\beta}(y) J^{\mu}(x) J_{5\alpha}(0)
               \right]\!
          |0
  \rangle\,,
\label{eq:Corr.JJJ}
\end{equation}
where $q$ corresponds to the photon momentum ($q^2=-Q^2$)
and $p_2$ is the outgoing pion momentum,
was described in detail in~\cite{IS82,NR82} for the case of local
condensates and in~\cite{BR91} for the non-local condensate (NLC) case.
In this correlator,
      $J^{\mu}(x)=   e_u\,\overline{u}(x)\gamma^\mu u(x)
                   + e_d\,\overline{d}(x)\gamma^\mu d(x)$
is the electromagnetic current, while
 $J_{5\alpha}(x)=   \overline{d}(x)\gamma_5\gamma_\alpha u(x)$ and
 $J^{+}_{5\beta}(x)=   \overline{u}(x)\gamma_5\gamma_\beta d(x)$
are axial-vector currents, where $e_u=2/3$ and $e_d=-1/3$ are the
electric charges of the $u$ and $d$ quarks, respectively.
Herewith one obtains the following SR:
\begin{eqnarray}
\label{eq:ffQCDSR}
  f_{\pi}^2\,F_{\pi}(Q^2)
=
  \int\limits_{0}^{s_0}\!\!\int\limits_{0}^{s_0}\!ds_1\,ds_2\
           \rho_3(s_1, s_2, Q^2)\,
           e^{-(s_1+s_2)/M^2}
     + \Phi_\text{G}(Q^2,M^2)
     + \Phi_{\langle\bar{q}q\rangle}(Q^2,M^2)\,,
\end{eqnarray}
where the quark condensate contribution
\begin{eqnarray}
\nonumber
\Phi_{\langle\bar{q}q\rangle}(Q^2,M^2)
  =
     \Phi_\text{4Q}(Q^2,M^2)
    +\Phi_\text{2V}(Q^2,M^2)
    +\Phi_{\bar qAq}(Q^2,M^2)
\end{eqnarray}
contains the four-quark condensate (4Q), the bilocal vector-quark
condensate (2V), and the antiquark-gluon-quark condensate term
($\bar q Aq$).
The gluon-condensate contribution to SR (\ref{eq:ffQCDSR}) is presented
by the $\Phi_\text{G}(Q^2,M^2)$ term.
The graphical illustration of the corresponding diagrams
is shown in Fig.\ \ref{fig:JJJ.NLO}.
The perturbative three-point spectral density entering the SR above
reads
\begin{eqnarray}
 \label{eq:SpDen.pert}
  \rho^{(1)}_3(s_1, s_2, Q^2)
  &=& \left[\rho_3^{(0)}(s_1, s_2, Q^2)
        + \frac{\alpha_s(Q^2)}{4\pi}\,
           \Delta\rho_3^{(1)}(s_1, s_2, Q^2)
    \right]\,.
\end{eqnarray}
Recall that the leading-order spectral density
\begin{eqnarray}\label{eq:RoSq}
\rho_3^{(0)}(s_1, s_2, t)
    &=&
       \frac{3}{4\pi^2}
       \left[t^2 \frac{d^2}{dt^2}
             + \frac{t^3}{3} \frac{d^3}{dt^3}
       \right]
       \frac{1}{\sqrt{\left(s_1 + s_2 + t\right)^2 - 4\,s_1s_2}}\
\end{eqnarray}
has been calculated in the early eighties~\cite{IS82,NR82},
whereas the explicit (but too complicated to show it here) expression
of the analogous next-to-leading order (NLO) version
$\Delta\rho_3^{(1)} (s_1, s_2, Q^2)$
has been obtained quite recently in~\cite{BO04}.
Note that the contribution from higher-resonances (HR), $F_\text{HR}$,
is usually modeled with the help of the same spectral density
\begin{equation}
 \label{eq:HR}
  \rho_\text{HR}(s_1, s_2)
  =  \left[1-\theta(s_1<s_0)\theta(s_2<s_0)\right]\,
      \rho_3(s_1, s_2, Q^2)
\end{equation}
and using the continuum threshold parameter $s_0$.

A novelty of the present investigation is the use of a running
coupling that shows by construction analytic behavior in $Q^2$,
i.e., has no Landau singularities.
This is done by adopting the arguments and techniques used in our
previous works in~\cite{BRS00,BPSS04}.
To be specific, we will use the one-loop analytic
Shirkov--Solovtsov coupling~\cite{SS97}
\begin{eqnarray}
 \label{eq:alphaS}
  \alpha_s(Q^2)
   &=&\frac{4\pi}{b_0}
       \left(\frac{1}{\ln(Q^2/\Lambda_{\text{QCD}}^2)}
           - \frac{\Lambda_{\text{QCD}}^2}{Q^2-\Lambda_{\text{QCD}}^2}
                \right)\,,
\end{eqnarray}
with $b_0=9$ and $\Lambda_\text{QCD}=300$~MeV.
The interested reader can find more about this subject in the
recent reviews \cite{SS06,AB08,Ste09}.
The nonperturbative terms
$\Phi_\text{G}$ and $\Phi_{\langle\bar{q}q\rangle}$
in the local-condensate case are well-known~\cite{NR82,IS82}:
\begin{eqnarray}
 \label{eq:Phi.G.qq.Loc}
 \Phi_\text{G}^\text{loc}(M^2)
  = \frac{\langle\alpha_s GG\rangle}{12\,\pi\,M^2}
 \,,~~~~~~
 \Phi_{\langle\bar{q}q\rangle}^\text{loc}(Q^2,M^2)
  = \frac{104\,A_0}{M^4}\,
     \left(1+\frac{2\,Q^2}{13\,M^2}\right)\,.
\end{eqnarray}
The standard QCD SRs~\cite{NR82,IS82} for the pion form factor
are based on these local expressions which show a wrong scale behavior
at large $Q^2$.
This means that the quark contribution contains both a linearly growing
term as well as a constant one, while the gluon contribution
is just a constant (see Table \ref{tab:History}).
At the same time, the first, i.e., the perturbative, term on the
right-hand side of Eq.\ (\ref{eq:ffQCDSR}) behaves at large $Q^2$
as $s_0/Q^4$ or $M^2/Q^4$.
For this reason, the SR becomes unstable for $Q^2>3$~GeV$^2$.
Hence, in order to be able to extract predictions from a SR of this
sort in the $Q^2$ region between $3-10$~GeV$^2$, we have first to
improve the quality of the SR.

\begin{table}[hb]\vspace*{-3mm}
\caption{
 $Q^2$-behavior of the nonperturbative contribution in different QCD SR
 approaches.
 Here $c_1,~c_2,~c_3,~c_4$ are dimensionless constants (not depending
 on $Q^2$).
 The abbreviations used are: LD for local duality, LO for leading
 order, and NLO for next-to-leading order, while $\lambda_q^2$ and
 $M^2$ denote the vacuum quark nonlocality parameter
 and the Borel parameter,
 respectively.
\label{tab:History}\vspace*{+1mm}}
\begin{ruledtabular}
\begin{tabular}{lccc} 
{\strut$\vphantom{\vbox to 6mm{}}\vphantom{_{\vbox to 4mm{}}}$}
Approach                                &Accuracy  &Condensates
& $Q^2$-behavior of $\Phi_{\text{OPE}}$                   \\ \hline
{\strut$\vphantom{\vbox to 6mm{}}\vphantom{_{\vbox to 4mm{}}}$}
Standard QCD SR~\cite{NR82,IS82}        &LO        &local
& $c_1 + Q^2/M^2$                                         \\ \hline
{\strut$\vphantom{\vbox to 6mm{}}\vphantom{_{\vbox to 4mm{}}}$}
QCD SR with NLCs~\cite{BR91}            &LO        &local $+$ nonlocal
& $\left(c_2+Q^2/M^2\right)\left(e^{- c_3 Q^2\lambda_q^2/M^4}+c_4\right)$
                                                          \\ \hline
{\strut$\vphantom{\vbox to 6mm{}}\vphantom{_{\vbox to 4mm{}}}$}
LD QCD SR~\cite{NR84,BO04,BLM07}        &NLO       &no
& $0$                                                     \\ \hline  
{\strut$\vphantom{\vbox to 6mm{}}\vphantom{_{\vbox to 4mm{}}}$}
Here                                    &NLO       &nonlocal
& $\left(c_1 + Q^2/M^2\right)\,e^{- c_3 Q^2\lambda_q^2/M^4}$       
\end{tabular}
\end{ruledtabular}
\end{table}

The fact that the condensate contributions in the SR for the pion form
factor are constant, or even growing with $Q^2$, is somewhat
surprising because exactly the corresponding diagrams should actually
generate \emph{decreasing} contributions as $Q^2$ increases.
But recall that the diagrams generating the condensate contributions
differ from the ordinary Feynman diagrams of QCD perturbation theory.
They result from the replacement of some of the propagators by constant
factors which represent condensates.
For example, the quark propagator
$\langle T(q(z)\bar q(0))\rangle$ is substituted by the quark
condensate $\langle \bar q(0)q(0)\rangle$.
As a result, instead of obtaining a $Q^2$-dependent contribution, one
gets a constant one.
The dependence on $Q^2$ appears when one calculates the contributions
of higher-dimension operators of the type
$\langle \bar q(0)D^2q(0)\rangle$,
$\langle \bar q(0)(D^2)^2q(0)\rangle$
etc., entailed by the Taylor expansion of the original nonlocal
condensate (NLC), where $\langle \bar q(0)q(z)\rangle$
is the nonperturbative part of the quark propagator.
The resulting total condensate contribution decreases for large $Q^2$.
However, each term of the standard OPE has the structure $(Q^2/M^2)^n$,
and one should, therefore, resum them to get a meaningful result.
Our strategy is to avoid the original Taylor expansion and deal
instead directly with the NLC.
This leads to a modified diagram technique involving new lines and
vertices that correspond to the NLC.
Then, the simplest contribution to $\Phi_{\langle\bar{q}q\rangle}$
is due to the vector condensate $M_{\mu}$ (cf.\ (\ref{eq:M_i})); viz.,
\begin{eqnarray}
 \Delta\Phi_V(M^2, Q^2)
  = \frac{8\,A_0}{M^4}\,
     \left(2+\frac{Q^2}{M^2-\lambda_q^2}\right)
      \exp\left[\frac{-\,Q^2\,\lambda_q^2}
                     {2\,M^2\left(M^2-\lambda_q^2\right)}
          \right]\,.
\end{eqnarray}
Obviously, this term indeed vanishes for large $Q^2$, as expected.
Moreover, the larger the nonlocality parameter $\lambda_q^2$, the
faster this contribution decreases with $Q^2$.
The value $Q^2_{*}$ at which this decrease starts, strongly
depends on the value of $M^2$.
Adopting for the nonlocality $\lambda_q^2=0.4$~GeV$^2$, one
finds for $M^2=1, 1.5,2$~GeV$^2$,
$Q^2_{*}=2, 6, 13$~GeV$^2$, respectively.

\begin{figure}[hb]
 \centerline{\includegraphics[width=0.33\textwidth]{
  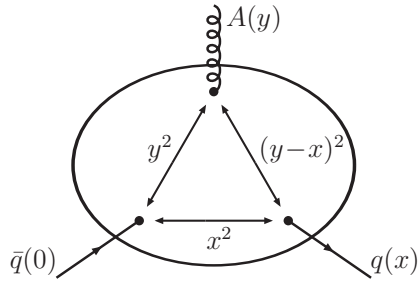}}
   \caption{\label{fig:3L}
    Illustration of the distance dependence of the three-point
    condensate in terms of the mutual inter-parton separations.}
\end{figure}
A first attempt \cite{BR91} to generalize the QCD SR to the NLC case
employed the specific model (\ref{eq:fi.barqAq.BR91})
for the three-point quark-gluon-quark NLC (with the NLC
$M_i(x^2,y^2,(x-y)^2)$ given by Eq.\ (\ref{eq:M_i})) and assuming a
nonlocality only with respect to two inter-parton separations, say,
$x^2$ and $(x-y)^2$,
out of the three possible separations $x^2$, $y^2$, and $(x-y)^2$
(see Fig.\ \ref{fig:3L}).
Hence, this approach is only partially nonlocal.
The following parametric functions
($\Lambda=\lambda_q^2/2$) were used:
\begin{eqnarray}
\label{eq:fi.barqAq.BR91}
 f^\text{BR}_i\left(\alpha,\beta,\gamma\right)
&=&
         \delta\left(\alpha-x_{i1}\Lambda\right)
         \delta\left(\beta -x_{i2}\Lambda\right)
         \delta\left(\gamma-x_{i3}\Lambda\right)\,,
\\\nonumber
 x_{ij}
&=&
   \left(
        \begin{array}{ccc}
           0.4 & 0   & 0.4\\
           0   & 1   & 0.4\\
           0   & 0.4 & 0.4
        \end{array}
  \right)\,.
\end{eqnarray}
The zero elements in the matrix $x_{ij}$ indicate the absence of
nonlocality effects either for the quark-antiquark separation $y^2$
($i=1, j=2$)
or for the antiquark-gluon separation $x^2$ ($i=2, 3$ and $j=1$)---see
Fig. \ref{fig:3L}.
Therefore, also this approach partially suffers from the same
shortcomings as the standard QCD SR.

Fortunately, the NLC contributions to the pion form factor taken into
account in the SR of \cite{BR91}, have a model-independent form,
allowing us to use them in connection with more improved versions
of the quark-gluon NLC.
To be more specific, we apply here QCD SR employing the minimal
(\ref{eq:Min.Anz.qGq}) \cite{MR92, BM98,BMS01} and the improved
(\ref{eq:Imp.Anz.qGq}) \cite{BP06} Gaussian models of NLC.
The corresponding parameters of the NLC are listed
in Appendix~\ref{App:A.0}.
An advantage of employing NLCs is that their use considerably
enlarges the region of applicability of the QCD SR to momenta as
high as $10~\gev{2}$.
In addition, we extend the accuracy of the perturbative spectral
density to the NLO level by including contributions of
$O(\alpha_s)$ \cite{BO04}---in contrast to previous works
\cite{BR91,IS82,NR82} in which only a LO perturbative spectral
density was taken into account.
It turns out that the NLO contribution influences the prediction for
the pion form factor, calculated with the described method,
reaching the level of $20\%$.

\section{Pion form factor analysis}
 \label{sec:pion-FF-analysis}
\subsection{Calculation and predictions}
 \label{subsec:details}

The strategy to further process the obtained SR is standard:
At each fixed value of $Q^2$,
SR\ (\ref{eq:ffQCDSR}) gives us the pion form factor
$F_{\pi}(Q^2,M^2,s_0)$ as a function of two additional parameters,
notably, the auxiliary Borel parameter $M^2$ and the continuum
threshold $s_0$.
The parameter $s_0$ can be interpreted as the boundary between the pion
and the higher resonances ($A_1$, $\pi'$, etc.).
We assume that $s_0$ should not be lower than the middle point,
0.6~GeV$^2$, of the interval between $m^2_\pi=0$ and
$m^2_{A_1}\approx 1.6$~GeV$^2$.
\begin{table}[b]\vspace*{-3mm}
\caption{
 Boundaries of the fiducial intervals $M_{\pm}^{2}/2$ with
 $M^2\in[M^2_{-}/2, M^2_{+}/2]$
 and values of the pion-decay constant ($f_{\pi}$) used in the
 two-point sum rules with nonlocal condensates in connection with the
 minimal and the improved Gaussian model for the nonlocality.
 \label{tab:CIandfpi}\vspace*{+1mm}}
\begin{ruledtabular}
\begin{tabular}{l|cccl}
 Model               & $f_\pi$         & $M^2_{-}$      & $M^2_{+}$
 \\ \hline
 Minimal~\cite{BMS01}& $0.137$~GeV$^2$ & $1$~GeV$^2$ & $1.7$~GeV$^2$\\
 Improved~\cite{BP06}& $0.142$~GeV$^2$ & $1$~GeV$^2$ & $1.9$~GeV$^2$
\end{tabular}
\end{ruledtabular}
\end{table}

\begin{figure}[t]
 \centerline{\includegraphics[width=0.47\textwidth]{
  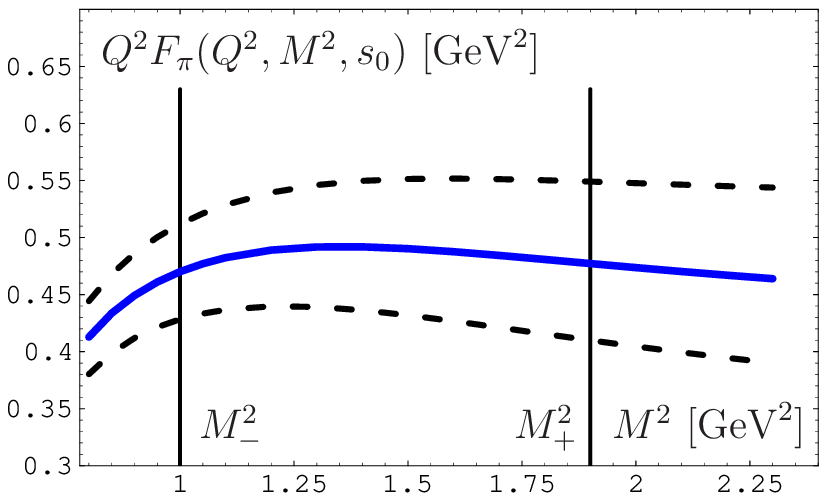}~~~%
             \includegraphics[width=0.47\textwidth]{
  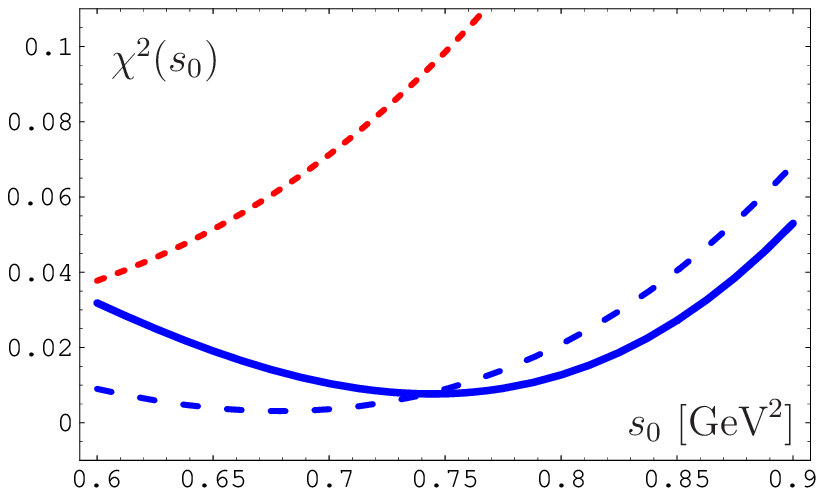}}
   \caption{\label{fig:Q2FF(M2).Xi(s0)}\footnotesize
    Left panel: Dependence of the pion form factor
    $Q^2 F_{\pi}(Q^2)$ at $Q^2=5$~GeV$^2$
    on the auxiliary Borel parameter $M^2$ in the improved NLC model.
    The solid blue line corresponds to $s_0=0.75$~GeV$^{2}$,
    whereas the dashed lines refer to $s_0=0.65$~GeV$^{2}$
    (upper curve) and $s_0=0.85$~GeV$^{2}$ (lower curve).
    Right panel: Root-mean-square deviation $\chi^2(Q^2,s_0)$ for the
    minimal NLC model at $Q^2=3$~GeV$^{2}$ (long-dashed blue line) and
    at $Q^2=5$~GeV$^{2}$ (short-dashed red line).
    The improved model of NLC is shown as a solid blue line at
    $Q^2=5$~GeV$^{2}$.}
\end{figure}

The specific value of $s_0(Q^2)$ at a given value of $Q^2$ is
determined by the condition implied by the minimal sensitivity of the
function $F_{\pi}(M^2,s_0)$ on the auxiliary parameter $M^2$ in the
fiducial interval of the SR.
We derive these intervals and the values of the pion decay constant
$f_{\pi}$ from the corresponding two-point NLC QCD SR, employing both
the minimal and the improved model for the vacuum nonlocality---see
Table \ref{tab:CIandfpi} for details.
Note here that the value of the Borel parameter $M^2$ in the
three-point SR roughly corresponds to the Borel parameter in the
two-point SR, having, however, twice its magnitude:
$M_\text{three-point}^2=2M_\text{two-point}^2$,
as given in Table \ref{tab:CIandfpi}.
In the left panel of Fig. \ref{fig:Q2FF(M2).Xi(s0)}, we show how
the scaled pion form factor $Q^2\,F_{\pi}(Q^2, M^2, s_0)$
depends on $M^2$ for three different values of the threshold parameter:
$s_0=0.65$, $0.75$, and $0.85$~GeV$^2$ at $Q^2=5$~GeV$^2$.
As a rule, the higher the value of $s_0$, the larger the
form factor because the perturbative input increases.

Using the root-mean-square deviation $\chi^2(Q^2,s_0)$,
given by Eq.\ (\ref{eq:xi}),
we determine that continuum threshold $s_0^{\text{SR}}(Q^2)$
which minimizes the dependence of the right-hand side
of (\ref{eq:ffQCDSR}) on the Borel parameter
$M^2\in[M^2_{-}, M^2_{+}]$
at each value of $Q^2$.
For an illustration, we refer to the right panel of
 Fig.\ \ref{fig:Q2FF(M2).Xi(s0)}.

As one sees from the right panel of Fig.\ \ref{fig:Q2FF(M2).Xi(s0)}
the long-dashed line---which corresponds to the minimal NLC model at
$Q^2=5$~GeV$^2$---has no minimum in the relevant $s_0$ interval.
Therefore, we cannot extract a reliable continuum-threshold value
$s_0^{\text{SR}}(Q^2\geq4~\text{GeV}^2)$ for this model using the
root-mean-square deviation criterion.

\begin{figure}[b]
 \centerline{\includegraphics[width=0.50\textwidth]{
  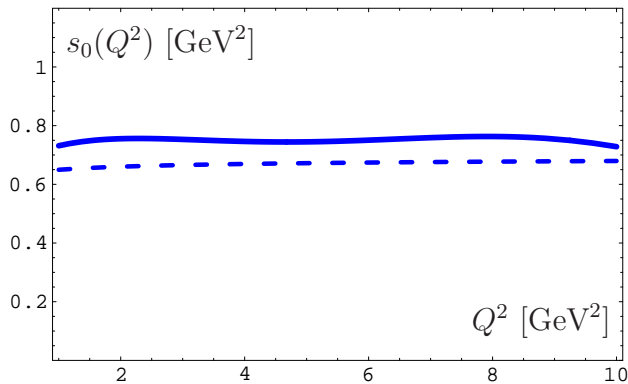}
             }
   \caption{\label{fig:s0(M2).Q2FF}\footnotesize
   Continuum threshold $s_0(Q^2)~[\gev{2}]$ for the minimal
   (dashed line) and for the improved (solid line) NLC model.
             }
\end{figure}
Note, however, that the values $\min\limits_{s}[\chi^2(Q^2,s)]$
and $\chi^2(Q^2,s_0=s_0^{\text{LD};(1)}(Q^2)\simeq0.63~\text{GeV}^2)$
are very close to each other, the relative difference for
$Q^2=4-10$~GeV$^2$ being of the order of $10-15$\%\footnote{%
Here $s_0^{\text{LD};(1)}(Q^2)$ is the standard Local-Duality
prescription for the continuum threshold, see the discussion after
Eq.\ (\ref{eq:LD.s0}).}.
For this reason, we will use in the minimal model
$s_0^{\text{SR}}(Q^2)=s_0^{\text{LD};(1)}(Q^2)$
as the continuum threshold.
In contrast, in the improved case (the solid line in the right panel
of  Fig.\ \ref{fig:Q2FF(M2).Xi(s0)}), the root-mean-square deviation
has a minimum approximately at the same point
$s_0^{\text{SR}}(Q^2)\approx 0.75$~GeV$^2$
for any $Q^2$ value within the considered interval.
The continuum thresholds $s_0^{\text{SR}}(Q^2)$ for the minimal
(dashed line) and the improved (solid line) NLC model, obtained this
way, are shown in Fig.~\ref{fig:s0(M2).Q2FF}.

\begin{figure}[b]
 \centerline{\includegraphics[width=0.60\textwidth]{
  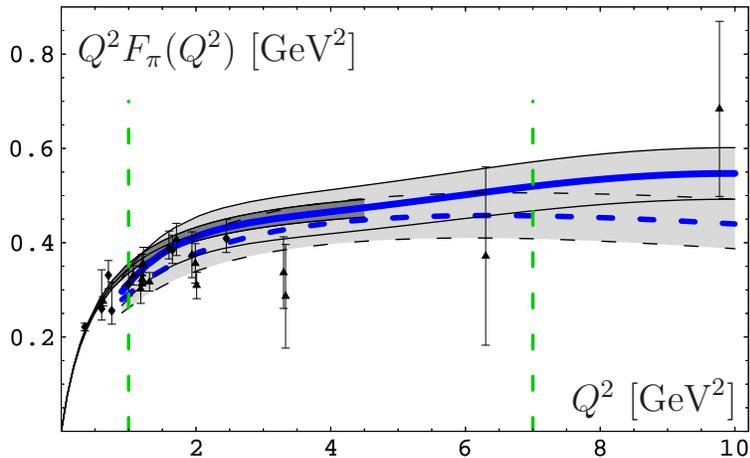}}
   \caption{\label{fig:main-result}\footnotesize
   Scaled pion form factor $Q^2 F_{\pi}(Q^2)$ for the minimal (dashed
   blue lines) and for the improved (solid blue lines) NLC model using
   $\lambda_q^2=0.4~\gev{2}$ in comparison with experimental data of
   the Cornell \protect\cite{FFPI74,FFPI76,FFPI78} (triangles)
   and the JLab Collaboration \cite{JLab08II} (diamonds).
   The shaded band delimited by the dashed lines corresponds to
   the minimal model, whereas the analogous band bounded by solid
   lines represents the uncertainty range of the improved model.
   The recent lattice result of \cite{Brommel06} is shown as a monopole
   fit with error bars between the two thick lines at lower $Q^2$.
   The two broken vertical lines denote the region, where the influence
   of the particular Gaussian model used to parameterize the QCD vacuum
   structure in the NLC QCD sum rules is not so strong.
   }
\end{figure}

\begin{figure}[b]
 \centerline{\includegraphics[width=0.49\textwidth]{
  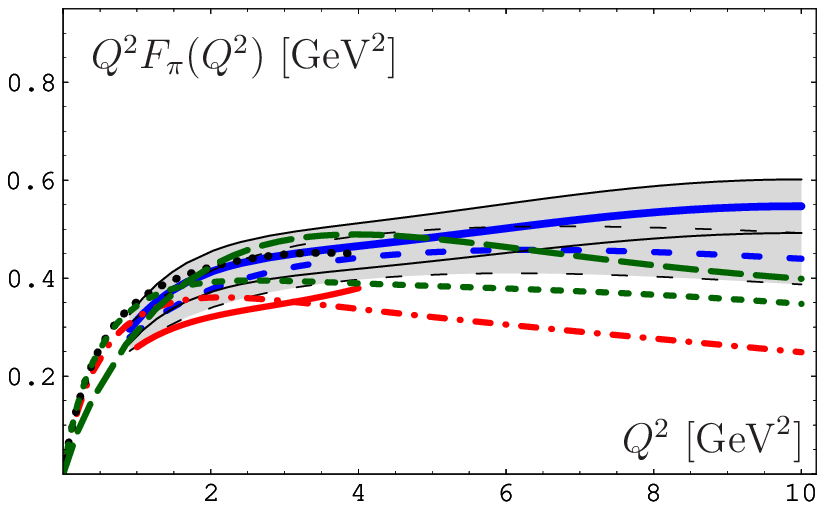}~~~
             \includegraphics[width=0.49\textwidth]{
  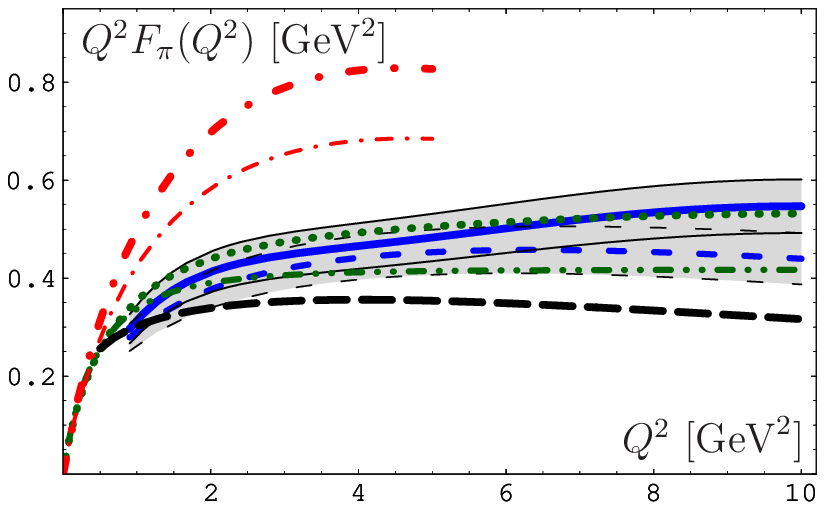}}
   \caption{\label{fig:Q2FFall}\footnotesize
   Comparison of theoretical predictions for the scaled pion form
   factor $Q^2 F_{\pi}(Q^2)$ obtained with a variety of theoretical
   methods and models.
   Our estimates are shown in both panels as shaded bands corresponding
   to the minimal NLC model (dashed blue lines) and the improved NLC
   model (solid blue lines).
   In both cases
   $\lambda_q^2=0.4~\gev{2}$ has been used and the associated error
   range is indicated by the width of the shaded bands.
   Left panel:
   The first short-dashed green line above to the dash-dotted red one
   is from \cite{RAB08} and is based on a large-$N_c$ Regge model.
   The long-dashed green line is from \cite{RA09}, obtained by
   including radiative and higher-twist effects within the framework
   of resummed pQCD.
   The short solid red line at low $Q^2$ shows the result from the
   standard QCD SR with local condensates \cite{NR82,IS82}.
   The dash-dotted red line denotes the estimate from \cite{BLM07}
   which was derived from Local Duality QCD SR.
   The dotted line represents the model of \cite{MT00}, based on the
   Bethe--Salpeter equation.
   Right panel:
   The heavy long-dashed line shows the result obtained with an
   AdS/QCD pion DA in \cite{AN08}.
   The dash-dot-dotted green line shows a Hirn--Sanz-type
   holographic model result, discussed in \cite{GR08}.
   The dotted green line gives the prediction of the AdS/QCD
   soft-wall model \cite{BT07}.
   The heavy and the fine dash-dotted red lines show, respectively,
   the predictions derived from the improved soft and hard wall AdS/QCD
   background approach~\cite{KL08}.
   }
\end{figure}

On that basis, we can define the SR result for the pion form factor as
the average value of the right-hand side of (\ref{eq:ffQCDSR}) with
respect to the Borel parameter $M^2\in[M^2_{-}, M^2_{+}]$:
\begin{eqnarray}
 \label{eq:FF.pi.SR.result}
 F_{\pi}^\text{SR}(Q^2)
  &=& \frac{1}{M_{+}^2-M_{-}^2}
       \int_{M_{-}^2}^{M_{+}^2}
        F(Q^2, M^2, s_0^\text{SR}(Q^2))\,dM^2\,.
\end{eqnarray}
Our main predictions for the minimal and the improved NLC model,
using in both cases $\lambda_q^2=0.4~\gev{2}$,
are shown in Fig.\ \ref{fig:main-result} in the form of two bands,
each one corresponding to the particular NLC model used, with the
width of each band denoting the inherent theoretical uncertainties
of the underlying QCD SR method.
The band within the dashed lines contains the predictions for the
minimal model.
Its counterpart for the improved model is limited by the solid
lines.
For the central curves (dashed---minimal model; solid---improved model)
in Fig.\ \ref{fig:main-result}, as well as in both panels of
Fig.\ \ref{fig:Q2FFall}, we used the following interpolation formulas:
\begin{subequations}
\begin{eqnarray}
 \label{eq:FF.pi.SR.Int.Min}
  F_{\pi;\text{min}}^{\text{SR}}(Q^2=x~\text{GeV}^2)
   &=& e^{-1.402\,x^{0.525}}
        \left(1+0.182\,x+\frac{0.0219\,x^3}{1+x}
        \right)\,,\\
 \label{eq:FF.pi.SR.Int.Imp}
  F_{\pi;\text{imp}}^{\text{SR}}(Q^2=x~\text{GeV}^2)
   &=& e^{-1.171\,x^{0.536}}
        \left(1+0.0306\,x+\frac{0.0194\,x^3}{1+x}\right)\,,
\end{eqnarray}
\end{subequations}
valid for $Q^2\in[1,10]$~GeV$^2$, i.~e., for $x\in[1,10]$.
The two broken vertical lines in Fig.\ \ref{fig:main-result}
denote the strict fidelity window of the NLC QCD sum rules.
We have limited this window from above by the requirement that the
predictions obtained with both NLC models have an overlap.
As one sees, the central curve for the minimal model starts departing
from the band of the improved model just around $7$~GeV$^2$.
But the predictions remain useful (though less accurate) even at higher
$Q^2$ values close to 10~GeV$^2$.
Our results compare favorably with the lattice calculation
of \cite{Brommel06},
shown as a monopole fit with associated
error bars between the two thick lines at lower $Q^2$.
We observe a similarly good agreement with the existing experimental
data of the Cornell \cite{FFPI74,FFPI76,FFPI78} (triangles)
and the JLab Collaboration \cite{JLab08II} (diamonds).

Figure \ref{fig:Q2FFall} shows our results in comparison with
the predictions obtained with various theoretical models (left
panel), whereas the right panel compares our results with
predictions from AdS/QCD models.
Specifically, in the left panel of Fig.\ \ref{fig:Q2FFall}
we display as a short-dashed green line the prediction
from \cite{RAB08},
whereas the long-dashed green line is from \cite{RA09}.
The short heavy solid red line at low $Q^2$ represents
the standard QCD SR result
with local condensates \cite{NR82,IS82},
while the dash-dotted red line gives the more recent estimate
of the Local Duality QCD SR of \cite{BLM07}.
Note here that the results of~\cite{BPSS04}, which are not shown in
this figure, are approximately $20$\% higher than those represented
by the dash-dotted red curve.
This is due to (i) the inclusion of the $O(\alpha_s^2)$ correction
(10\%) and (ii) the uncertainty of the matching procedure, used in
\cite{BPSS04}, (10\%) (see the discussion in Sect.\
\ref{subsec:LD-approach} and the graphics in Fig.\
\ref{fig:Q2FFLDvsMOD}).
Finally, the dotted line below 4~GeV$^2$ displays the result
extracted from the model of \cite{MT00} which uses the Bethe--Salpeter
equation.

In the right panel of Fig.\ \ref{fig:Q2FFall} we collect recent
estimates from the following AdS/QCD models:
The heavy long-dashed line displays the result obtained with an
AdS/QCD pion DA \cite{AN08}.
The dash-dot-dotted green line represents a Hirn--Sanz-type
holographic model, discussed in \cite{GR08}.
The dotted green line provides the prediction of the AdS/QCD soft model
\cite{BT07}.
Finally, the two top broken red lines show the results obtained from
the improved soft-wall (heavy dash-dotted line) and the hard-wall
(fine dash-dotted line) AdS/QCD background approach~\cite{KL08},
respectively.
One sees that the dotted and the dash-dot-dotted green lines are in
compliance with our predictions,
favoring the results of~\cite{GR08},
as these leave some space to add radiative corrections.

\begin{figure}[t]
 \centerline{\includegraphics[width=0.49\textwidth]{
  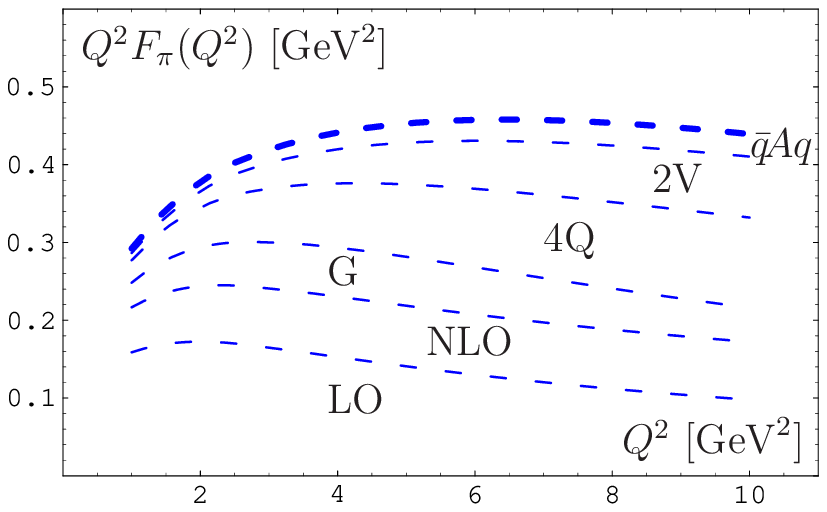}~~~~
             \includegraphics[width=0.49\textwidth]{
  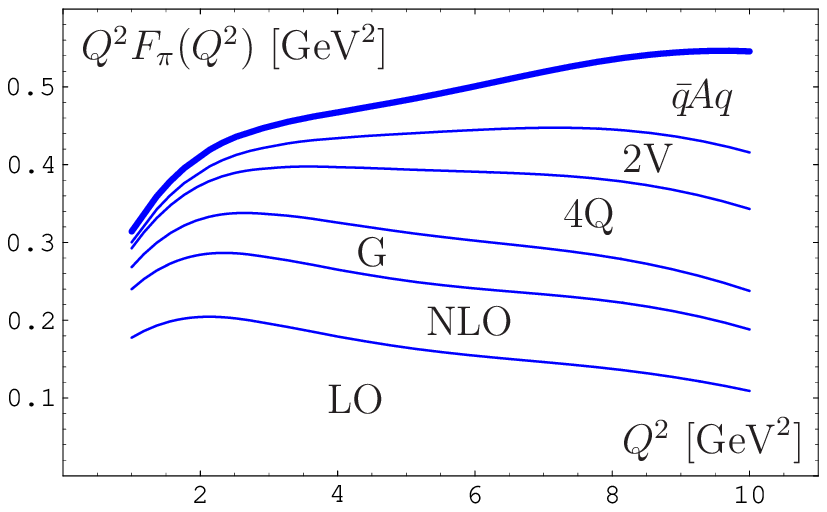}}
   \caption{\label{fig:Q2FFapart}\footnotesize
        Contributions to the pion form factor from the
        $O\left(1\right)$ and
        $O\left(\alpha_s\right)$ spectral-density terms (LO and NLO).
        The specific contributions are denoted by labels referring to
        their particular origin: 4-quarks (scalar) condensate---(4Q),
        bilocal vector quark condensate---(2V),
        tri-local quark-gluon-quark condensate---($\bar{q}Aq$),
        and gluon condensate---(G)
        in the minimal (left panel) and the improved (right panel)
        model.}
   \end{figure}

Figure \ref{fig:Q2FFapart} serves to illustrate
the origin of the differences
between the minimal (dashed line---left panel) and
the improved (solid line---right panel) Gaussian NLC model
with respect to their corresponding predictions
for the pion form factor.
The difference between them can be traced back to the quark-gluon-quark
contribution $\Phi_{\bar qAq}(Q^2,M^2)$ which is different for each
model (top lines in both panels below the final results).
This, in turn, has influence on the continuum threshold $s_0(Q^2)$
entailing changes in the corresponding LO and NLO terms, albeit mild
ones.
This figure exhibits in detail how the various contributions to the
pion form factor are accumulated.
Each curve from bottom to top is the sum of all previous terms, while
the heavy lines on the top denote the full result for each model.

Note that the calculations were made for a nonlocality parameter
$\lambda_q^2=0.4~\gev{2}$, a value receiving support from a recent
comprehensive analysis \cite{BMS02,BMS03} of the CLEO data on the
pion-photon transition.
Using higher values of this parameter, would entail a decrease of the
pion form factor owing to a stronger influence of the nonlocality
effects.
Evidently, using smaller values of $\lambda_q^2$ would cause
an increase of the pion form factor.
The predictions shown in our figures, discussed above, provide
further support for a value of $\lambda_q^2$ in the range
compatible with the result extracted from the CLEO-data analysis.

\subsection{Local duality approach}
 \label{subsec:LD-approach}
Sum rules based on local duality have no condensate contributions due
to the $M^2\rightarrow \infty$ limit.
On the other hand, the determination of the threshold $s_0$ is not
possible using this method, in contrast to QCD SR.
One could, nevertheless, try to extract it from (local duality) sum
rules for the pion-decay constant $f_\pi$, but this would correspond to
a value of the pion form factor at a small $Q^2$ value
because of the Ward identity~\cite{BLM07}.

As originally argued in Refs. \cite{NR82,IS82,NR84,IL84,Rad90}, the
dominant contribution to the pion form factor at low up to moderate
values of the momentum transfer $Q^2 \leq 10~\text{GeV}^2$
originates mainly from the soft part that involves no hard-gluon
exchanges but may be attributed to the Feynman mechanism.
Using the Local Duality (LD) approach to calculate the soft contribution,
it is assumed that the pion form factor is dual to the free quark
spectral density~\cite{NR82,Rad95}.
Then, staying within the $(l+1)$-loop order\footnote{%
The triangle diagram at the one-loop level does not include any
radiative corrections from the very beginning.
For this reason, the $O(\alpha_s^l)$-correction for this diagram
appears first at the $(l+1)$-loop order.}, one has
\begin{subequations}
 \label{eq:FF.LD}
  \begin{eqnarray}
   F_{\pi}^{\text{LD};(l)}(Q^2)
    &=& F_{\pi}^{\text{LD};(l)}(Q^2,s_0^{\text{LD};(l)}(Q^2))\, ,\\
   F_{\pi}^{\text{LD};(l)}(Q^2,S)
    &\equiv& \frac{1}{f_{\pi}^2}
      \int_{0}^{S}\!\!\!\!\int_{0}^{S}\!\!\!
       \rho_3^{(l)}(s_1, s_2, Q^2)\,ds_1\,ds_2\,,
\end{eqnarray}
\end{subequations}
where $s_0^{\text{LD};(l)}(Q^2)$ is the LD effective threshold
parameter for the higher states in the axial channel and the
three-point $(l+1)$-loop spectral density is
$\rho_{3}^{(l)}(s_1, s_2, Q^2)$.
In leading order,
we know $\rho_{3}^{(0)}(s_1, s_2, Q^2)$ from Eq.\ (\ref{eq:RoSq}),
so that
\begin{eqnarray}
 F_{\pi}^{\text{LD};(0)}(Q^2,S)
  &=& \frac{S}{4\pi^2f_\pi^2}\,
       \left(1 - \frac{Q^2 + 6 S}
                      {Q^2 + 4 S}\,
                  \sqrt{\frac{Q^2}{Q^2 + 4 S}}
       \right)\,.
\label{eq:FF.LD.LO}
\end{eqnarray}
The LD prescription for the corresponding correlator \cite{SVZ,Rad95}
implies the relations
\begin{eqnarray}
 s_0^{\text{LD};(0)}(0)
  = 4\,\pi^2\,f_{\pi}^2
 \quad\text{~and~}\quad
 s_0^{\text{LD};(1)}(0)
  = \frac{4\,\pi^2\,f_{\pi}^2}
         {1+\alpha_s(Q_{0}^2)/\pi}\,,
 \label{eq:LD.s0}
\end{eqnarray}
where $Q_{0}^2$ is of the order of $s_0^{\text{LD};(0)}(0)$.
This prescription is a strict consequence of the Ward identity
for the AAV correlator due to the vector-current conservation.
In principle, the $Q^2$ dependence of the LD parameter
$s_0^{\text{LD}}(Q^2)$ (\ref{eq:FF.LD})
should be determined from the QCD SR at $Q^2\gtrsim1$~GeV$^2$.
As we have already explained in Sec.~\ref{sec:vac}, the standard
QCD SR becomes unstable at $Q^2>3$~GeV$^2$ because of the appearance
of terms in the condensate contributions linearly growing with $Q^2$
\cite{IS83,NR84}.
For this reason, this dependence was known only for
$Q^2\leq 3$~GeV$^2$ and, therefore, most authors usually used the
constant approximation
$s_0^{\text{LD};(0)}(Q^2)\simeq s_0^{\text{LD};(0)}(0)$,
like in~\cite{NR82,BRS00,BPSS04,BO04}, or a slightly $Q^2$-dependent
approximation
$ s_0^{\text{LD};(1)}(Q^2)
\simeq
 4\,\pi^2\,f_{\pi}^2/(1+\alpha_s(Q^2)/\pi)
$,
like in~\cite{BLM07}.

\begin{figure}[ht]
 \centerline{\includegraphics[width=0.47\textwidth]{
  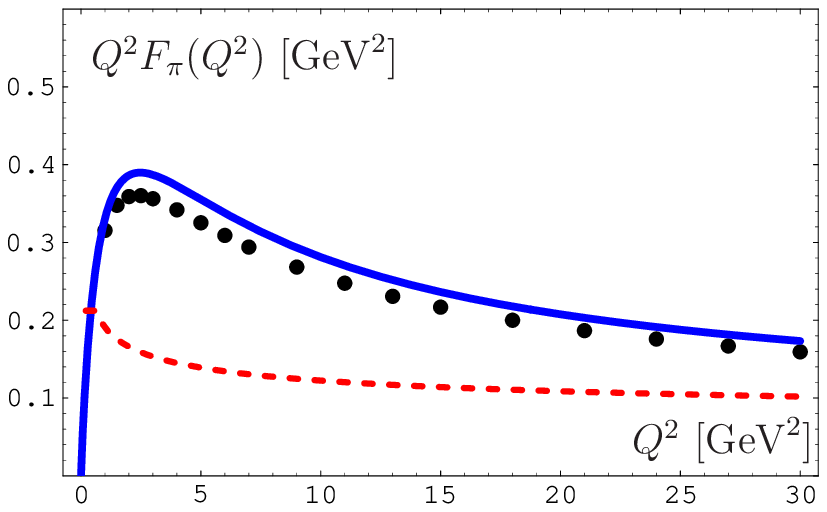}~~~
             \includegraphics[width=0.47\textwidth]{
  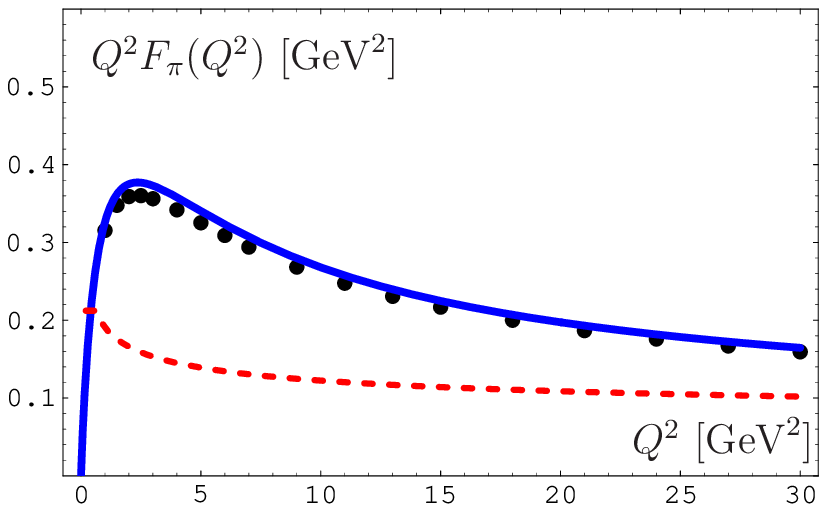}}
   \caption{\label{fig:Q2FFLDvsMOD}
    Comparison of the exact NLO LD result for the pion form factor
    $F_{\pi}^{\text{LD},(1)}(Q^{2})$ (black dots)
    with the models $F_{\pi}^{\text{WI};(1)}(Q^{2})$
    ((\ref{eq:Fpi-Mod.NLO}), left panel) and
    $F_{\pi;\text{imp}}^{\text{WI};(1)}(Q^{2},s_0^{\text{LD};(1)}(Q^2))$
    ((\ref{eq:Fpi-Mod.imp}), right panel),
    displayed in both panels as solid blue lines.
    The asymptotic perturbative prediction
    $F_{\pi}^{\text{pQCD},(1)}(Q^{2})$
    is also shown as a dashed red line.}
\end{figure}

Lacking profound knowledge about the exact
structure of the NLO spectral density
$\rho_{3}^{(1)}(s_1,s_2,Q^2)$, two of us (A.B. and N.G.S.) with
collaborators have suggested in~\cite{BPSS04} to use for
the full pion form factor
the information about the factorizable part,
$F_{\pi}^{\text{pQCD},(2)}(Q^2)$, which is
computable within perturbative QCD (pQCD) (here evaluated at the
two-loop order).
But it should be noted that the pQCD term has the wrong limit
at $Q^2=0$, calling for a correction of this behavior
in order to maintain the Ward identity (WI) $F_{\pi}(0)=1$.
To achieve this goal, we apply a matching procedure, introduced in
\cite{BPSS04}, namely,
\begin{eqnarray}
 \label{eq:Fpi-Mod.NNLO}
  F_{\pi}^{\text{WI};(2)}(Q^{2})
   &=& F_{\pi}^{\text{LD},(0)}(Q^{2})
     + \left(\frac{Q^2}{2s_0^{(2)}+Q^2}\right)^2
        F_{\pi}^{\text{pQCD},(2)}(Q^2)
\end{eqnarray}
with $s_0^{(2)}\simeq0.6$~GeV$^2$.
This approximation was used to `glue' together the LD model for the
soft part,
$F_{\pi}^{\text{LD},(0)}(Q^{2})$
(which is dominant at small $Q^2\leq 1$~GeV$^2$),
with the perturbative hard-rescattering part,
$F_{\pi}^{\text{pQCD},(2)}(Q^2)$
(which provides the leading perturbative
$O(\alpha_s)+O(\alpha_s^2)$ corrections
and is dominant at large $Q^2\gg1$~GeV$^2$),
in such a way as to ensure the validity of the Ward identity
$F_{\pi}^{\text{WI};(2)}(0)=1$.
In order to test the quality of the matching prescription given by
Eq.\ (\ref{eq:Fpi-Mod.NNLO}), we propose to compare it with the LD
model (\ref{eq:FF.LD}) evaluated at the one-loop order
(i.e., in the $O(\alpha_s)$-approximation~\cite{BO04,BLM07}).
To this end, we construct the analogous $O(\alpha_s)$-model
\begin{subequations}
\begin{eqnarray}
 \label{eq:Fpi-Mod.NLO}
  F_{\pi}^{\text{WI};(1)}(Q^{2})
  &=&  F_{\pi}^{\text{LD},(0)}(Q^{2})
  + \frac{\alpha_s(Q^2)}{\pi}
      \frac{2\,Q^2\,s_0^{\text{LD};(0)}(0)}
      {(2\,s_0^{\text{LD};(1)}(Q^2)+Q^2)^2}\,,
\end{eqnarray}
where we made use of the asymptotic form factor \cite{CZS77,FJ79}
\begin{eqnarray}
 \label{eq:Fpi-Pert.NLO}
  F^{\text{pQCD},(1)}_\pi(Q^2)
  = \frac{8\,\pi\,f_\pi^2\,\alpha_s(Q^2)}{Q^2}
  = \frac{\alpha_s(Q^2)}{\pi}\,
        \frac{2\,s_0^{\text{LD};(0)}(0)}{Q^2} \, ,
\end{eqnarray}
implying the same prescription for the effective LD threshold
as in~\cite{BLM07}:
\begin{eqnarray}
 \label{eq:LD.s0.(1)}
  s_0^{\text{LD};(1)}(Q^2)
   &=& \frac{4\,\pi^2\,f_{\pi}^2}
            {1+\alpha_s(Q^2)/\pi}\,.
\end{eqnarray}
\end{subequations}
It is worth noting that the model $F_{\pi}^{\text{WI};(1)}(Q^{2})$,
following from the matching procedure (\ref{eq:Fpi-Mod.NLO})
suggested in~\cite{BPSS04}, works quite well, albeit it was proposed
without the knowledge of the exact two-loop spectral density, which
became available only somewhat later \cite{BO04}.
Recall that the key feature of the matching prescription is that it
uses information on $F_{\pi}(Q^2)$ in two asymptotic regions:
\begin{enumerate}
 \item $Q^2\to0$,
   where the Ward identity dictates $F_{\pi}(0)=1$
   and
   hence $F_{\pi}(Q^2)\simeq F_{\pi}^{\text{LD},(0)}(Q^2)$,
 \item $Q^2\to\infty$,
   where $F_{\pi}(Q^2)\simeq F_{\pi}^{\text{pQCD},(1)}(Q^2)$
\end{enumerate}
in order to join properly the hard tail of the pion form factor with
its soft part.
Numerical analysis of (\ref{eq:Fpi-Mod.NLO}) shows that the applied
prescription yields a pretty accurate result, with a relative error
varying in the range 5\% at $Q^2=1$~GeV$^2$ to 9\% at
$Q^2=3-30$~GeV$^2$.
The graphical comparison of $F_{\pi}^{\text{WI};(1)}(Q^{2})$
(solid blue line)
with the LD result (\ref{eq:FF.LD}) (black dots),
employing the two-loop spectral density~\cite{BO04},
is displayed in the left panel of Fig.~\ref{fig:Q2FFLDvsMOD}.
This figure also shows the purely perturbative part
$Q^2F_{\pi}^{\text{pQCD},(1)}(Q^{2})$
(dashed red line)---not corrected
by the factor $[Q^2/(2s_0+Q^2)]^2$ to the right
low-$Q^2$ behavior.\footnote{For this reason, this expression tends to
the finite value 0.21 GeV$^2$ at $Q^2\to0$ and does not vanish.}
[In similar context a matching procedure, as that described above, was
also used in a related work \cite{RA09} on the electromagnetic pion and
kaon form factor including radiative and higher-twist effects.]

Once the spectral density $\rho_{3}^{(1)}(s_1, s_2, Q^2)$
was calculated~\cite{BO04}, it made it possible to improve the
representation of the LD part in (\ref{eq:Fpi-Mod.NLO}) by taking
into account the leading $O(\alpha_s)$ correction in the
electromagnetic vertex.
On that basis, we suggest the following improved WI model
for
$F_{\pi;\text{imp}}^{\text{WI};(1)}(Q^{2},s_0^{\text{LD};(1)}(Q^2))$:
\begin{eqnarray}
  F_{\pi;\text{imp}}^{\text{WI};(1)}(Q^{2},S)
   &=& F_{\pi}^{\text{LD};(0)}(Q^{2},S)
    \nonumber\\
   &+& \frac{S}{4\pi^2f_\pi^2}\,
        \bigg\{\frac{\alpha_s(Q^2)}{\pi}\,
                \left(\frac{2S}{2S+Q^2}\right)^2
             + F^{\text{pQCD},(1)}_\pi(Q^2)\,
               \left(\frac{Q^2}{2S+Q^2}\right)^2
        \bigg\}\, .~~~
 \label{eq:Fpi-Mod.imp}
\end{eqnarray}

We explicitly display the dependence on the threshold $S$ in
Eq.\ (\ref{eq:Fpi-Mod.imp})---the aim being to apply it later on
with $S=s_0^{\text{LD-eff}}(Q^2)$,
the latter value being extracted by comparing the NLC QCD SR results
with the LD approximation.
Note that this model has the right limit at $Q^2\to0$,
provided one sets
$S=s_0^{\text{LD};(1)}(Q^2)=4\pi^2f_\pi^2/(1+\alpha_s(Q^2)/\pi)$.
Indeed, one has $F_{\pi}^{\text{LD};(0)}(0,S)=S/(4\pi^2f_\pi^2)$
by virtue of the fact that the term
$F^{\text{pQCD},(1)}_\pi(Q^2)$ is canceled
by the factor $[Q^2/(2S+Q^2)]^2$ in this limit,
so that the net result is\footnote{%
One usually applies some ``freezing'' assumption
$s_0^{\text{LD};(1)}(Q^2) = 4\pi^2f_\pi^2/(1+\alpha_s(s_0)/\pi)$ for
$Q^2\leq s_0\simeq0.6$~GeV$^2$.
In this case, the same ``freezing'' should be used
in Eq.\ (\ref{eq:Fpi-Mod.imp})
for the argument of $\alpha_s$ as well:
$\alpha_s(Q^2)\to\alpha_s(s_0)$
for $Q^2\leq s_0$.
Note that in order to treat the electromagnetic radius
of the pion (i.e., the derivative of the pion form factor
in the low $Q^2$ domain) correctly, one has to apply
a different form of the OPE and the corresponding SR---see
for details in~\cite{NR84}.}
$$F_{\pi;\text{imp}}^{\text{WI};(1)}(0,s_0^{\text{LD};(1)}(0))
= \frac{s_0^{\text{LD};(1)}(0)}{4\pi^2f_\pi^2}
  \left[1+\frac{\alpha_s(s_0)}{\pi}\right]
= 1.
$$
The graphical evaluation of this new WI model in comparison with the
exact LD result in the one-loop approximation is displayed
in the right panel of Fig.~\ref{fig:Q2FFLDvsMOD}.
We can see from this graphics that the quality of the matching
condition, given by (\ref{eq:Fpi-Mod.NLO}), is improved.
Indeed, the relative error is reduced, varying between
4\% at $Q^2=1-10$~GeV$^2$ and 3\% at $Q^2=30$~GeV$^2$.

Proceeding along similar lines of reasoning, we construct
the two-loop WI model
$F_{\pi}^{\text{WI};(2)}(Q^{2},s_0^{\text{LD};(2)}(Q^2))$
for the pion form factor to obtain
\begin{eqnarray}
  F_{\pi}^{\text{WI};(2)}(Q^{2},S)
   &=& F_{\pi}^{\text{LD};(0)}(Q^{2},S)
      \nonumber\\
   &+& \frac{S}{4\pi^2f_\pi^2}\,
        \bigg\{\frac{\alpha_s(Q^2)}{\pi}\,
                \left(\frac{2S}{2S+Q^2}\right)^2
             + F^{\text{FAPT},(2)}_\pi(Q^2)\,
               \left(\frac{Q^2}{2S+Q^2}\right)^2
        \bigg\}\,,~~~
 \label{eq:Fpi.WI-Mod.NNLO}
\end{eqnarray}
where $F^{\text{FAPT},(2)}_\pi(Q^2)$ is the analyticized
expression generated from $F^{\text{pQCD},(2)}_\pi(Q^2)$
using Fractional Analytic Perturbation Theory (FAPT)
(see Refs.\ \cite{KS01,Ste02,BMS05,BKS05,BMS06} and,
 in particular, \cite{AB08,AB08quarks})
to get a result which appears to be very close
to the outcome of the default scale setting
($\muR=\muF=Q^2$),
investigated in detail in \cite{BPSS04}.
The explicit expression for $F^{\text{FAPT},(2)}_\pi(Q^2)$ is given
in Appendix \ref{App:FAPT}.
This model provides the possibility to implement the perturbative QCD
$O(\alpha_s^2)$-results for the pion form factor
without performing an explicit three-loop calculation
of the three-point spectral density,
a welcome advantage, as this calculation is very tedious.
Moreover, the case of the one-loop approximation with the WI model
(\ref{eq:Fpi-Mod.imp}) indicates that the relative error of this
procedure is of the order of 10\%.
This level of accuracy allows us to estimate the weight of the
$O(\alpha_s^2)$-correction.
We find that its relative contribution to the pion form factor
is of the order of 10\%, as has been estimated in~\cite{BPSS04,AB08}.
Hence, the relative error of our estimate is of the order of
1\%---provided we take into account the $O(\alpha_s)$-correction
exactly via the specific choice of $s_0(Q^2)$,
as done in (\ref{eq:FF.LD.s0}).
In order to use this formula, we only need to improve our knowledge
about the effective LD thresholds $s_0^{\text{LD};(1)}(Q^2)$.

\begin{figure}[t]
 \centerline{\includegraphics[width=0.45\textwidth]{
  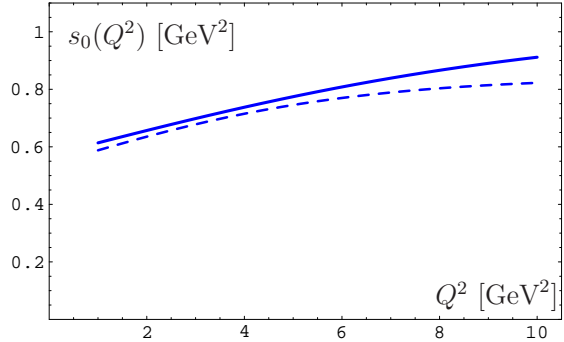}}
   \caption{\label{fig:s0LD}\footnotesize
    Effective continuum thresholds
     $s_{0,\text{imp}}^\text{LD-eff}(Q^2)$ (solid blue line) and
     $s_{0,\text{min}}^\text{LD-eff}(Q^2)$ (dashed blue line)
     that approximate the NLC QCD SR results using
     the LD $O(\alpha_s(Q^2))$-formulas.}
\end{figure}

As we also stated at the beginning of this section,
the problem of adjusting the continuum threshold parameter
$s_0^{\text{LD}}(Q^2)$ is of high importance for the LD approach.
From our point of view,
it should be determined by comparing the LD results with those
derived via the Borel QCD SRs.
In the previous section, we processed our NLC QCD SR and
obtained the interpolation expressions
(\ref{eq:FF.pi.SR.Int.Min}), (\ref{eq:FF.pi.SR.Int.Imp})
which are valid for $Q^2\in[1,10]$~GeV$^2$.
Now we can determine the corresponding effective thresholds
$s_{0,\text{min}}^\text{LD-eff}(Q^2)$ and
$s_{0,\text{imp}}^\text{LD-eff}(Q^2)$
in the minimal and the improved NLC model, respectively, to find
\begin{eqnarray}
 \label{eq:FF.LD.s0}
  F_{\pi;\text{imp}}^{\text{WI};(1)}
  \left(Q^2,s_{0,\text{min}}^\text{LD-eff}(Q^2)\right)
  &=& F_{\pi;\text{min}}^{\text{SR}}(Q^2)\, ,
  ~~~
  F_{\pi;\text{imp}}^{\text{WI};(1)}
  \left(Q^2,s_{0,\text{imp}}^\text{LD-eff}(Q^2)\right)
 \ =\ F_{\pi;\text{imp}}^{\text{SR}}(Q^2)\,.~~~~
\end{eqnarray}
The solutions of these equations, namely,
$s_{0,\text{min}}^\text{LD-eff}(Q^2)$
and $s_{0,\text{imp}}^\text{LD-eff}(Q^2)$, are shown in
Fig. \ref{fig:s0LD}; they can be represented in this range of
$Q^2$ by the following interpolation formulas:
\begin{subequations}
\begin{eqnarray}
 \label{eq:FF.LD.s0.App}
  s_{0,\text{min}}^\text{LD-eff}(Q^2=x~\text{GeV}^2)
  &=& 0.57+0.307\,\tanh(0.165\,x)
          -0.0323\,\tanh(775\,x)\, ,\\
  s_{0,\text{imp}}^\text{LD-eff}(Q^2=x~\text{GeV}^2)
  &=& 0.57+0.461\,\tanh(0.0954\,x)\,.~~~
\end{eqnarray}
\end{subequations}
We see that both thresholds turn out to increase only moderately with
$Q^2$.

\begin{figure}[h]
 \centerline{\includegraphics[width=0.47\textwidth]{
  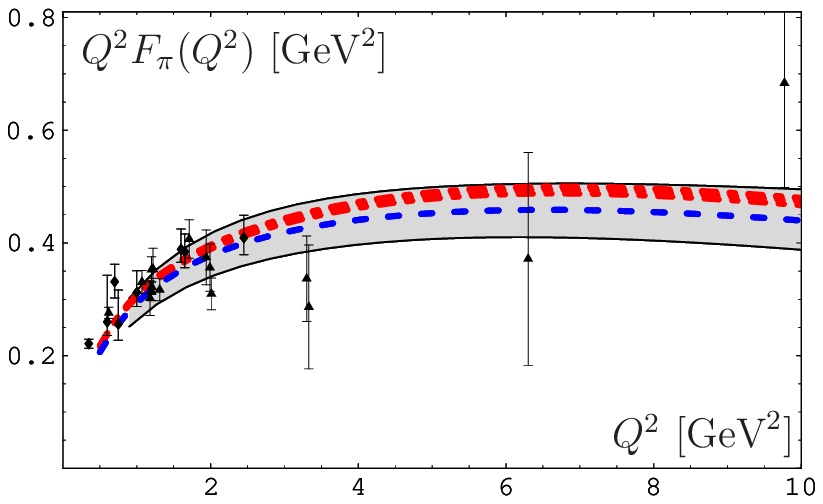}~~~
             \includegraphics[width=0.47\textwidth]{
  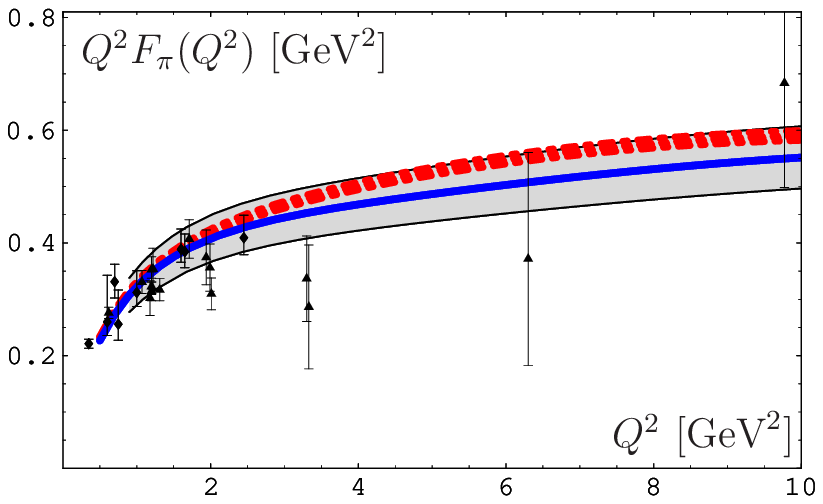}}
     \caption{\label{fig:FF.LD.2Loop}\footnotesize
     We show as a narrow dash-dotted red strip the predictions
     for the pion form factor, obtained in the two-loop WI model,
     Eq.\ (\ref{eq:Fpi.WI-Mod.NNLO}), using the minimal (left panel)
     and the improved (right panel) Gaussian model.
     The width of the strip is due to the variation of the Gegenbauer
     coefficients $a_2$ and $a_4$ (needed to calculate the collinear
     part $F_{\pi}^{\text{pQCD},(2)}(Q^2)$)
     in the corresponding shaded bands for the pion DA in
     both models (indicated by the central blue lines).
     The normalization of the strip is also affected by the
     effective continuum thresholds
     $s_{0,\text{min}}^\text{LD-eff}(Q^2)$ (left panel)
     and $s_{0,\text{imp}}^\text{LD-eff}(Q^2)$ (right panel).
}
\end{figure}

The results obtained for the pion form factor with our two-loop
model, i.e.,
Eq.\ (\ref{eq:Fpi.WI-Mod.NNLO}),
and using the effective LD thresholds
$s_{0,\text{min}}^\text{LD-eff}(Q^2)$ and
$s_{0,\text{imp}}^\text{LD-eff}(Q^2)$,
are displayed in Fig.\ \ref{fig:FF.LD.2Loop}.
We see from this figure
that the main effect of the NNLO correction
peaks at $Q^2\gtrsim4$~GeV$^2$, reaching the level of $3-10$\%,
and can be estimated by taking recourse to the results of the FAPT
analysis in \cite{BPSS04,AB08}.

\subsection{Estimation of uncertainties}
 \label{subsec:uncertainties}

Even with our methodologically improved approach,
there are still intrinsic uncertainties
that influence our results.
Therefore we consider in this subsection two potential sources of
uncertainties:
(i) the choice of $\Lambda_\text{QCD}$ and
(ii) those uncertainties originating from our nonperturbative
assumptions.
\begin{itemize}
\item
We performed our numerical estimates using a value of
$\Lambda_\text{QCD}=300$~MeV for 3 flavors of active quarks.
Let us here discuss how our results change as we vary the value of
this parameter.
First, we note that the relative contribution of the NLO correction to
the value of the pion form factor is of the order of 20\%,
as we already mentioned at the end of Section~\ref{sec:vac}.
Next, the relative difference between
$\alpha_s(Q^2,\Lambda_\text{QCD}=300~\text{MeV})$ and
$\alpha_s(Q^2,\Lambda_\text{QCD}=200~\text{MeV})$
is of the order of 15\% for $Q^2=1-10$~GeV$^2$.
Hence, varying the QCD scale parameter $\Lambda_\text{QCD}$ in the
range $200-300$~MeV induces uncertainties of our predictions of the
order of about $3\%$.
\item
As it was shown in \cite{BP06}, the minimal Gaussian model for the
nonlocal quark condensate does not satisfy the QCD equations of motion
and the transversality condition for the two-point correlator of the
vector currents.
For this reason, using the minimal model can induce some artificial
term in the OPE part of the QCD SR.
Bearing this in mind, the improved model was constructed in such a way
as to minimize this unphysical contribution for the first five
lowest-order conformal moments of the pion DA.
Speaking in terms of a Taylor expansion of the nonlocal
condensates, this improvement pertains to the correction of the
low-dimensional local condensates---which represent coefficients in
this expansion.
The dark side of this improvement is that it could render the
definition of higher-dimensional local condensates ill-defined
relative to the minimal model.
This is the reason why we refrained from applying the improved
model to the extraction of the higher conformal moments of the pion DA
and the high-momentum behavior of the pion form factor.
As we can see from Fig.\ \ref{fig:Q2FFapart}, in the case of the
improved model the quark-gluon term ($\bar q A q$) grows with
$Q^2$ in the considered momentum region---in contrast
to the minimal model.
We conclude from this that the $\bar q A q$-term in the improved
model is not well-defined in the high $Q^2$ region.
One can realize from Fig.\ \ref{fig:Q2FFall} that the difference
between the minimal and the improved model tends to become larger
relative to the accuracy of the SR at $Q^2\approx 7$~GeV$^2$.
This means that the validity of the nonperturbative model used
is lost at this point.
Therefore, we should anticipate a reduced accuracy of the obtained
results on the pion form factor in the high-momentum regime.
To overcome this problem, one should eventually set stricter
constraints on the model using experimental data---a promising option
after the planned upgrade of the CEBAF accelerator.
Still another possibility comes from theory, notably, current
conservation and the equations of motion.
Such attempts are currently under scrutiny.
\end{itemize}

\section{Discussion and Conclusions}
\label{sec:Conclusions}

In this work we have calculated the pion's electromagnetic form
factor using a three-point QCD sum rule with nonlocal condensates.
The main scope of our investigation was on the methodological side,
though the obtained predictions compare well with both the existing
experimental data and lattice simulations.
The applied method offers the following key advantages:
\begin{enumerate}
  \item[(a)]
  The profile of the pion distribution amplitude is irrelevant.
  This removes a serious source of theoretical uncertainty
  entailed at the normalization point of the order of 1~GeV and
  simplifies the computation because the technically demanding task of
  the NLO evolution of the pion DA to higher $Q^2$ values to compare
  with data is absent.
  \item[(b)]
  Because this type of approach tends to become unstable above
  $Q^2>3$~GeV$^2$, we had to improve the quality of the sum rules
  considerably.
  This was achieved by dealing not with a Taylor expansion but
  directly with the nonlocal condensates, avoiding any local
  contribution.
  As a result, we were able to enlarge the region of applicability
  of the QCD SR towards momenta as high as $10~\text{GeV}^2$.
  \item[(c)]
  In order to minimize the transversality violations of the two-point
  correlator of vector currents, we considered in our analysis not
  only the minimal Gaussian model for the distribution of the quark
  average momentum in the vacuum, but also an improved version of it
  introduced earlier in \cite{BP06}.
  \item[(d)]
  A spectral density was used in the dispersion integral of the sum
  rules which includes terms of $O(\alpha_s)$.
  The influence of this next-to-leading-order (NLO) contribution to
  the pion form factor turns out to be quite important.
  \item[(e)]
  An analytic running coupling was used in the calculations pertaining
  to perturbation theory that helps surmount the problem of a Landau
  singularity at $Q^2=\Lambda_\text{QCD}^2$.
\end{enumerate}

The phenomenological findings of our analysis and their implications
are the following.
\begin{enumerate}
  \item[(i)]
  Our main predictions for $F_\pi(Q^2)$ were shown in
  Fig.\ \ref{fig:main-result} in comparison with the existing
  experimental data from the early Cornell \cite{FFPI74,FFPI76,FFPI78}
  and the more recent JLab Collaboration \cite{JLab08II}.
  We found that the $O(\alpha_s)$-contribution to the spectral density
  influences the pion form factor at the level of $20\%$.
  This estimate contrasts with the result obtained in
  \cite{BO04,BLM07}, which was found to be somewhat larger.
  \item[(ii)]
  Inspection of Fig.\ \ref{fig:main-result} reveals that the
  central-line curve of the improved model lies within the error range
  of the minimal model up to values $Q^2\approx 7$~GeV$^2$,
  indicating a comparable quality of both models in this momentum
  region.
  We pointed out that the main differences between these two models
  for the vacuum nonlocality can be attributed to the different
  contributions related to the quark-gluon-quark contribution
  $\Phi_{\bar qAq}(Q^2,M^2)$ (see Fig.\  \ref{fig:Q2FFapart}).
  \item[(iii)]
  In Fig.\ \ref{fig:Q2FFall} we compared our predictions with a
  variety of theoretical models for the electromagnetic pion
  form factor---including also results derived from holographic models
  based on the AdS/QCD correspondence.
  We found that the models of \cite{GR08} and \cite{BT07} are within
  the error bands of the minimal and the improved model, respectively.
  \item[(iv)]
  Applying a matching method of the factorized pion form factor,
  calculable within perturbative QCD, to its value at $Q^2=0$,
  subject to the Ward identity, we were able to model the pion form
  factor at the two-loop level without having recourse
  to the spectral density---appealing only to
  the local duality concept.
  This method was previously developed and applied successfully
  in \cite{BPSS04} and was used quite recently with an appropriate
  modification in \cite{RA09} in the calculation of the pion and the
  kaon form factors at the twist-three level.
  The key element in our approach is the extraction
   of the effective continuum thresholds $s_0^{\text{LD}}(Q^2)$
   from the results of the NLC QCD SR.
  \item[(v)]
  The results obtained with the local duality model, just described,
  were found to be inside the error band of the QCD sum rules with the
  nonlocal condensates (Fig.\ \ref{fig:FF.LD.2Loop}) providing support
  for the consistency of both approaches.
  Both methods yield predictions comparing well with the experimental
  data.
  \item[(vi)]
  Comparison of our predictions with those found with the LD approach
  \cite{BLM07} reveals that they are systematically higher than these.
  The reason for this difference is that the effective LD threshold
  $s_0^\text{LD}(Q^2)$ has a well-defined value only in the
  small-$Q^2$ region.
  For higher $Q^2$ values, it is not firmly fixed.
  The authors of \cite{BLM07} proposed therefore a logarithmically
  increasing threshold
  \begin{eqnarray*}
  s_0^\text{LD}(Q^2)
  =
  \frac{4\pi^2 f_\pi^2}{1+\alpha_s(Q^2)/\pi}\,,
  \end{eqnarray*}
  which is about
  $0.67$~GeV$^2$ at $Q^2\approx10$~GeV$^2$.
  In order to imitate the NLC QCD SR results within the LD approach,
  one needs to use
  $s_0^\text{LD}(Q^2=10~\text{GeV}^2)=0.87~\text{GeV}^2$.
  This means that the $s_0^\text{LD}$ uncertainty in the region of
  $Q^2=10$~GeV$^2$ is of the order of $20$\%.
\end{enumerate}

Bottom line:
 The use of three-point QCD sum rules with nonlocal condensates
 provides a reliable alternative to calculate the spacelike
 pion's electromagnetic form factor in that momentum region
 which is accessible to current and planned experiments.

\section*{Acknowledgements}
We are grateful to S.~V.~Mikhailov for helpful discussions.
Two of us (A.P.B. and A.V.P.) are indebted to Prof.\ Klaus Goeke
for the warm hospitality at Bochum University, where part of this
work was done.
The reported investigation was supported in part by the Deutsche
Forschungsgemeinschaft under contract DFG 436 RUS 113/881/0,
the Heisenberg--Landau Programme, grant 2009,
the DAAD Michail-Lomonosov-Forschungsstipendium,
the Russian Foundation for Fundamental Research,
grants No.\ ü~07-02-91557, 08-01-00686, and 09-02-01149,
and the BRFBR--JINR Cooperation Programme, contract No.\ F08D-001.

\begin{appendix}
\section{Parametrization of the nonlocal condensates}
 \renewcommand{\theequation}{\thesection.\arabic{equation}}
  \label{App:A.0}\setcounter{equation}{0}
We use, as usual in the QCD sum-rules approach, the fixed-point
(Fock--Schwinger) gauge $x^\mu A_\mu(x)=0$.
For this reason, all connectors
$
 {\mathcal C}(x,0)
\equiv
 {\mathcal P}
  \exp\!\left[-ig_s\!\!\int_0^x t^{a} A_\mu^{a}(y)dy^\mu\right]=1
$,
assuming for the integration contour a straight line going from $0$
to $x$.
For the scalar and the vector condensates, we apply the
same minimal model, as used in \cite{BM98,BMS01}:
\begin{eqnarray}
\label{eq:S.V.NLC}
 \langle{\bar{q}(0)q(z)}\rangle
=
 \langle{\bar{q}q}\rangle\,
     e^{-|z^2|\lambda_q^2/8}\,;\;
 \langle{\bar{q}(0)\gamma_\mu q(z)}\rangle
=
 \frac{i\, z_\mu\,z^2}{4}\,
    A_0\
      e^{-|z^2|\lambda_q^2/8}\,,
\end{eqnarray}
where $A_0=2\alpha_s\pi\langle{\bar qq}\rangle^2/81$.
The nonlocality parameter
$\lambda_q^2 = \langle{k^2}\rangle$
characterizes the average momentum of quarks in the QCD vacuum and
has been estimated in QCD SR~\cite{BI82,OPiv88} and on the
lattice~\cite{DDM99,BM02} to have a value around
$\lambda_q^2 = 0.45\pm 0.1\text{~GeV}^2$.
For the vector and the axial-vector quark-gluon-antiquark
condensate we use a parametrization suggested in~\cite{MR89,MR92}:
\begin{eqnarray}
\langle{\bar{q}(0)\gamma_\mu(-g\widehat{A}_\nu(y))q(x)}\rangle
   &=&
      (y_\mu x_\nu-g_{\mu\nu}(y\cdot x))\overline{M}_1(x^2,y^2,(y-x)^2)
      \nonumber\\
   &+&
      (y_\mu y_\nu-g_{\mu\nu}y^2)\overline{M}_2(x^2,y^2,(y-x)^2)\,,
      \nonumber
\\
\langle{\bar{q}(0)\gamma_5\gamma_\mu(-g\widehat{A}_\nu(y))q(x)}\rangle
    &=&
       i\varepsilon_{\mu\nu yx}\overline{M}_3(x^2,y^2,(y-x)^2)
\vspace{-5mm}\nonumber
\end{eqnarray}
with
\begin{eqnarray}
\label{eq:M_i}
 \overline{M}_i(x^2,y^2,z^2)
  = A_i\int\!\!\!\!\int\limits_{\!0}^{\,\infty}\!\!\!\!\int\!\!
        d\alpha \, d\beta \, d\gamma \,
         f_i(\alpha ,\beta ,\gamma )\,
          e^{\left(\alpha x^2+\beta y^2+\gamma z^2\right)/4}\,.
\end{eqnarray}
Note that here the following abbreviation
$A_{1,2,3}\equiv A_0 \times\left(-\frac32,2,\frac32\right)$
has been used.
The minimal model of the nonlocal QCD vacuum relies upon the following
ansatz:
\begin{eqnarray} \label{eq:Min.Anz.qGq}
f_i\left(\alpha,\beta,\gamma\right)
   = \delta\left(\alpha -\Lambda\right)\,
     \delta\left(\beta  -\Lambda\right)\,
     \delta\left(\gamma -\Lambda\right)
\end{eqnarray}
with $\Lambda=\lambda_q^2/2$.
By construction, this type of model faces problems with the QCD
equations of motion and the deficiency of the two-point correlator of
the vector currents to satisfy the transversality condition.
In order to fulfil the QCD equations of motion and at the same time
to minimize the non-transversality of the $VV$ correlator,
an improved model of the QCD vacuum was suggested in \cite{BP06}
that uses a modified Gaussian ansatz, viz.,
\begin{subequations}
\label{eq:Imp.Anz.qGq}
\begin{eqnarray}\label{eq:Imp.Anz.qGq.f}
f^\text{imp}_i\left(\alpha,\beta,\gamma\right)
  = \left(1 + X_{i}\partial_{x}
            + Y_{i}\partial_{y}
            + Y_{i}\partial_{z}
    \right)
         \delta\left(\alpha-x\Lambda\right)
         \delta\left(\beta -y\Lambda\right)
         \delta\left(\gamma-z\Lambda\right)\,,
\end{eqnarray}
where $z=y$, $\Lambda=\frac12\lambda_q^2$ and
\begin{eqnarray}
  X_1 &=& +0.082\,;~X_2 = -1.298\,;~X_3 = +1.775\,;~x=0.788\, ,~~~\\
  Y_1 &=& -2.243\,;~Y_2 = -0.239\,;~Y_3 = -3.166\,;~y=0.212\,.~~~
\end{eqnarray}
\end{subequations}
By virtue of the QCD equations of motion, these parameters satisfy
the supplementary conditions:
\begin{eqnarray}
 \label{eq:3L.D.A.Rule4}
  12\,\left(X_{2} + Y_{2}\right)
  - 9\,\left(X_{1} + Y_{1}\right)
  = 1 \,,~~~~~x+y=1\,.
\end{eqnarray}
On the other hand, the vacuum condensates of the four-quark operators
are reduced to a product of two scalar quark condensates
(\ref{eq:S.V.NLC}) by virtue of the vacuum-dominance hypothesis
\cite{SVZ}.

\section{QCD sum-rule parameters}
 \renewcommand{\theequation}{\thesection.\arabic{equation}}
  \label{App:A.1}\setcounter{equation}{0}
The parameters to characterize the nonlocal condensates
are
$\Lambda=\lambda_q^2/2=0.2$~GeV$^2$,
$\langle\alpha_s{GG}\rangle/\pi=0.012$~GeV$^4$
and
$\alpha_s\,\langle\bar{q}q\rangle^2$ = $1.83\cdot 10^{-4}$~GeV$^6$.
The nonlocal gluon-condensate contribution
$\Phi_\text{G}(M^2)$
produces a very complicated expression.
In analogy to the quark case, we model it by an exponential factor
\cite{BR91,MS93}:
$
 \Phi_\text{G}(M^2)
=
 \Phi_\text{G}^\text{loc}(M^2)\,e^{-\lambda_g^2 Q^2/M^4}
$
with $\lambda_g^2=0.4$~GeV$^2$.

In order to determine the best value of the threshold $s_0$,
we define a $\chi^2$ function for each value of $Q^2$ and $s_0$:
\begin{eqnarray}\label{eq:xi}
 \chi^2(Q^2,s_0)
  = \frac{\varepsilon^{-2}}{N_M}
    \left[\sum\limits_{i=0}^{N_M}
           Q^4\,F(Q^2,M^2_{i},s_0)^2
        - \frac{\left(\sum\limits_{i=0}^{N_M}
                       Q^2\,F(Q^2,M^2_{i},s_0)
                \right)^2}
               {N_M+1}
    \right]\,,
\end{eqnarray}
where we used $M_{i}^2=M_{-}^2+i\,\Delta_M$,
$\Delta_M=(M_{+}^2-M_{-}^2)/N_M$,
$N_M=20$,
and with $\varepsilon$ denoting the desired accuracy for
$\chi^2\simeq1$
(the actual value in the computation is $\varepsilon=0.07$~GeV$^2$.)

\section{Factorizable part of the pion form factor in Fractional
         Analytic Perturbation Theory}
 \renewcommand{\theequation}{\thesection.\arabic{equation}}
  \label{App:FAPT}\setcounter{equation}{0}
Here, we provide some technical details about the FAPT
analytization procedure in the computation of the factorizable part
of the pion form factor using perturbation theory.
In perturbative QCD at the NLO level and
with the scale setting $\muR=Q^2$ and $\muF=\text{const}$
(where the subscripts F and R denote, respectively,
 the factorization and the renormalization scale),
one obtains
\begin{eqnarray}
 F_{\pi}^{\text{pQCD},(2)}(Q^2;\muF)
  &=& \alpha_s^{(2)}(Q^2)\,
       \mathcal F_{\pi}^\text{LO}(q^2;\muF)
    + \frac{\left[\alpha_{s}^{(2)}(Q^2)\right]^2}{\pi}\,
       \mathcal F_{\pi}^\text{NLO}(Q^2;\muF) \ ,
 \label{eq:pff.pQCD(2)}
\end{eqnarray}
where the superscript $^{(2)}$ in $\alpha_s^{(2)}(Q^2)$
denotes the two-loop order of the coupling,
with
\begin{eqnarray}
 \label{eq:Q2pffLO}
  \mathcal F_{\pi}^\text{LO}(Q^2;\mu_\text{F}^2)
   &\equiv& \frac{8\,\pi\,f_{\pi}^2}{Q^2}\,
             \left[1 + a_2^\text{LO}(\mu_\text{F}^2)
                     + a_4^\text{LO}(\mu_\text{F}^2)
             \right]^2
\end{eqnarray}
and
\begin{subequations}
 \label{eq:Q2.pff.NLO}
\begin{eqnarray}
 {\mathcal F}_\pi^\text{NLO}(Q^2;\mu_\text{F}^2)
 & \! \equiv \! &
    b_0 {\mathcal F}_{\pi}^{(1,\beta)}(Q^2;\mu_\text{F}^2)
  + {\mathcal F}_{\pi}^{(1,\text{FG})}(Q^2;\mu_\text{F}^2)
  + C_\text{F}
    {\mathcal F}_\pi^{(1,\text{F})}(Q^2;\mu_\text{F}^2)
    \ln\left[\frac{Q^2}{\mu_\text{F}^2}\right] .~~~~~
 \label{eq:Q2.pff.NLO.Dia}
\end{eqnarray}
In Eq.\ (\ref{eq:pff.pQCD(2)}), and below,
$a_{2}^\text{LO}$ and $a_{4}^\text{LO}$
are the LO Gegenbauer coefficients
of the pion DA,
whereas the individual contributions in Eq.\ (\ref{eq:Q2.pff.NLO.Dia})
read (with further details and explanations given in
\cite{BPSS04,AB08})
\begin{eqnarray}
 {\mathcal F}_{\pi}^{(1,\beta)}(Q^2;\mu_\text{F}^2)
  = \frac{2\,\pi\,f_{\pi}^2}{Q^2}
    &&\!\!\!\!
      \left[\frac{5}{3}
         + \frac{3 + \displaystyle (43/6) a_2^\text{LO}(\mu_\text{F}^2)
                    + (136/15) a_4^\text{LO}(\mu_\text{F}^2)}
                      {1 + a_2^\text{LO}(\mu_\text{F}^2)
                    + a_4^\text{LO}(\mu_\text{F}^2)}
      \right]~~~~~\nonumber\\
    &&\!\!\!\!\times
    \Bigl[1+a_2^\text{LO}(\mu_\text{F}^2)
    +a_4^\text{LO}(\mu_\text{F}^2)\Bigr]^2\,,
 \label{eq:Q2.pff.NLO.beta}\\
 {\mathcal F}_{\pi}^{(1,\text{FG})}(Q^2;\mu_\text{F}^2)
  = -\frac{2\,\pi\,f_{\pi}^2}{Q^2}
  &&\!\!\!\!\Bigl\{ 15.67
                + a_2^\text{LO}(\mu_\text{F}^2)
                 \left[21.52 - 6.22\, a_2^\text{LO}(\mu_\text{F}^2)
                 \right]\nonumber\\
  &&+ a_4^\text{LO}(\mu_\text{F}^2)
                 \left[7.37 - 37.40\, a_2^\text{LO}(\mu_\text{F}^2)
                       - 33.61\, a_4^\text{LO}(\mu_\text{F}^2)
                 \right]
        \Bigr\} ,~~~~~
 \label{eq:Q2.pff.NLO.FG}\\
 {\mathcal F}_\pi^{(1,\text{F})}(Q^2;\mu_\text{F}^2)
  = -\frac{2\,\pi\,f_{\pi}^2}{Q^2}
  &&\!\!\!\!\!
        \left[\frac{25}{3}\,a_2^\text{LO}(\mu_\text{F}^2)
            + \frac{182}{15}\,a_4^\text{LO}(\mu_\text{F}^2)
        \right] \!
        \Bigl[1 + a_2^\text{LO}(\mu_\text{F}^2)
        + a_4^\text{LO}(\mu_\text{F}^2)
        \Bigr] .~~~~~~~
 \label{eq:Fcal.(1)F}
\end{eqnarray}
\end{subequations}
The analyticized NNLO contribution
(which includes radiative corrections of $O(\alpha_{s}^{2})$)
is then computed to be
\begin{eqnarray}
 \label{eq:piff.NNLO.APT}
  F_{\pi}^{\text{FAPT},(2)}(Q^2)
   &=& \mathcal A_{1}^{(2)}(Q^2)\,
        \mathcal F_{\pi}^\text{LO}(Q^2;\muF)
     + \frac{1}{\pi}\,
        \mathcal L_{2;1}^{(2)}(Q^2)\,
         \mathcal F_{\pi}^{(1,\text{F})}(Q^2;\muF)
  \nonumber\\
   &+& \frac{1}{\pi}\,
        \mathcal A_{2}^{(2)}(Q^2)\,
         \left[\mathcal F_{\pi}^\text{NLO}(Q^2;\muF)
             - \mathcal F_{\pi}^{(1,\text{F})}(Q^2;\muF)\,
                \ln\frac{Q^2}{\Lambda_\text{QCD}^2}
         \right]\,.
\end{eqnarray}
Here $\mathcal A_1(Q^2)$, $\mathcal A_2(Q^2)$, and
$\mathcal L_{2;1}(Q^2)$ are the analytic images
of $\alpha_s(Q^2)$, $\alpha^2_s(Q^2)$, and
$\alpha^2_s(Q^2)\ln(Q^2/\Lambda_\text{QCD}^2)$
\cite{BMS05,BKS05,AB08}:
\begin{eqnarray}
 \label{eq:An.A_1}
  \mathcal A_{1}(Q^2)
   &\equiv&
    \textbf{A}_\textbf{E}
     \left[\alpha_s(Q^2)\right]\, ,\\
 \label{eq:An.A_2}
  \mathcal A_{2}(Q^2)
   &\equiv&
    \textbf{A}_\textbf{E}
     \left[\alpha_s^{2}(Q^2)\right]\, ,\\
 \label{eq:An.L_21}
  \mathcal L_{2;1}(Q^2)
   &\equiv&
    \textbf{A}_\textbf{E}
     \left[\alpha_s^{2}(Q^2)\,
            \ln\left(\frac{Q^2}{\Lambda_\text{QCD}^2}\right)
     \right]
\end{eqnarray}
with
\begin{eqnarray}
 \textbf{A}_\text{E}\left[f(Q^2)\right]
  &\equiv&
   \int_0^{\infty}\!
    \frac{\textbf{Im}\,f(-\sigma)}
         {\sigma+Q^2}\,
       d\sigma\,.
\label{eq:A.E}
\end{eqnarray}
This type of approach was analyzed in \cite{BKS05} and it was shown
that the analytization procedure significantly reduces the dependence
of the results on the factorization-scale setting.
More recently \cite{AB08}, this procedure was further improved
by setting $\muF=Q^2$.
This renders the logarithmic term to cancel out, but swamps the
appearance of complicated expressions like
$
 \Ds\textbf{A}_\textbf{E}
 \left[\alpha_s^{p}(Q^2)\,a_{2n}^\text{LO}(Q^2)\,a_{2m}^\text{LO}(Q^2)
 \right]
$
with $p = 1, 2$ and $n, m = 0, 1, 2$.
Nevertheless, the obtained results are very close to those previously
found in \cite{BKS05} and in \cite{BPSS04} with $\muR=\muF=Q^2$.
The improved result $F_{\pi}^{\text{FAPT},(2)}(Q^2)$ was recently
obtained in \cite{AB08}.

\end{appendix}

\setcounter{section}{0}



\end{document}